\documentstyle[11pt,epsf,fleqn]{article}
\begin{document}
\title{The neutral Higgs boson masses of MSSM at one-loop level in explicit CP violation scenario}
\author{S.W. Ham$^{(1)}$, S.K. Oh$^{(1,2)}$, E.J. Yoo$^{(2)}$, C.M. Kim$^{(2)}$, D. Son$^{(1)}$ 
\\
\\
{\it $^{\rm (1)}$ Center for High Energy Physics, Kyungpook National University}\\
{\it Daegu 702-701, Korea} \\
{\it $^{\rm (2)}$ Department of Physics, Konkuk University, Seoul 143-701, Korea}
\\
\\
}
\date{}
\maketitle
\begin{abstract}
The neutral Higgs sector of the minimal supersymmetric standard model (MSSM) in explicit CP violation scenario is 
investigated at the one-loop level.
Within the context of the effective potential formalism, the masses of the neutral Higgs bosons are calculated at 
the one-loop level by taking into account the contributions of the following loops of ordinary particles and superpartners: 
top quark, the scalar top quarks, bottom quark, the scalar bottom quarks, tau lepton, the scalar tau leptons, $W$ boson, 
the charged Higgs boson, the charginos, $Z$ boson, the scalar and pseudoscalar Higgs bosons, and the neutralinos.
Our calculation is an improvement in the sense that both  the terms which are quartic in the electroweak coupling constants into account, 
and the pseudoscalar Higgs loop contribution are explicitly included. 
\end{abstract}
\vfil
\eject
\section{Introduction}

For new physics beyond the standard model (SM) [1], the enlargement of the Higgs sector is considered as one of 
indispensable ingredients, especially for supersymmetric models. 
Among various supersymmetric extensions of the SM, the minimal supersymmetric standard model (MSSM) is the simplest 
one which has just two Higgs doublets in its Higgs sector [2].
In principle, in the models with multiple Higgs doublets, the violation of CP symmetry may be accomplished either 
spontaneously or explicitly by the mixing between the scalar and pseudoscalar Higgs parts [3, 4, 5]. 
In practice, however, it is known that at the tree level the Higgs potential of the MSSM conserves CP symmetry, 
because the complex phase can always be eliminated by rotating the Higgs fields. Thus, neither explicit nor 
spontaneous CP violation can happen in the MSSM at the tree level.
Even at the one-loop level the scenario of spontaneous CP violation is excluded, because the radiatively corrected 
Higgs potentential of the MSSM leads to a very light Higgs boson which is unacceptable by the CERN $e^+ e^-$ LEP2 data.

The remaining possibility at the one-loop level for the MSSM is then the explicit CP violation scenario. 
A number of investigations have been devoted to examine the explicit CP violation in the MSSM at one-loop level [6]. 
In those investigations it is observed that the tree-level situation is considerably modified when radiative 
contributions are included.  
In particular, the one-loop MSSM Higgs potential with explicit CP violation is found to allow the lightest neutral Higgs boson to 
possess a mass above the expemental lower bound from the LEP2 data. 
Therefore, the general consensus is that the MSSM at the one-loop level can accommodate explicit CP violation.

Recently, Ibrahim and Nath [7] have investigated explicit CP violation scenario, paying their attention to the 
phenomenological implications of the non-trivial CP phase on the chargino sector. 
They also have computed the scalar-pseudoscalar mixings arising from the neutralino sector, and compared with those 
from the chargino sector [8]. 
Their calculations are quite exhaustive since almost all the relevant loops are included: top quark, the scalar top quark, 
bottom quark, the scalar bottom quark, tau lepton, the scalar tau lepton, $W$ boson, the charged Higgs boson, 
the charginos, $Z$ boson, the neutral Higgs bosons, and the neutralinos. 
They have reported that in the MSSM with explicit CP violation the neutralino exchange corrections to the mixings 
of the CP-even sector and the CP-odd sector are comparable to the chargino exchange corrections [8]. 

We would also investigate, in more detail, the Higgs sector of the MSSM with explicit CP violation at the one-loop level.
In our investigation, we take  explicitly the terms which are quartic in the electroweak coupling constants into account.
We regard it consistent to consider the quartic terms in the one-loop effective potential contributed by the scalar top 
quark, the scalar bottom quark, the scalar tau lepton, the chargino, and the neutralinos, because the Higgs self 
contributions are essentially induced by them.

Moreover, in our investigation, the pseudoscalar Higgs loop contribution is included, when the one-loop contributions to 
the neutral Higgs boson masses are evaluated, as well as all the loops of the relevant particles and superpartners.
The inclusion of the pseudoscalar Higgs contribution is not only reasonable but also necessary, because there is a 
non-trivial correlation between the ordinary particles and the corresponding superpartners in the one-loop effective potential.
In the neutral sector of the MSSM, there are four neutralinos as superpartners, and as ordinary particles there are $Z$ boson 
and three neutral Higgs bosons, one of them being the pseudoscalar Higgs boson. 
Thus, the radiative contribution of the pseudoscalar Higgs loop should simultaneously be included, whenever the 
loops of $Z$ boson and the two scalar Higgs bosons are taken into account, corresponding to the loops of four neutralinos. 
But for the pseudoscalar Higgs contribution, the mass matrix for the neutral Higgs bosons at the one-loop level might be 
inconsistent and would yield incomplete masses for them and mixing angles among them.
In order to obtain reliable results for the Higgs productions and their decays, it is quite necessary to include the pseudoscalar 
Higgs contribution.

In this paper, we evaluate the masses of the neutral Higgs bosons in the MSSM with explicit CP violation at the one-loop level.
We take into account the loops of the pseudoscalar Higgs boson as well as all the loops that have been considered in 
the investigations of Ref. [8], namely, the loops of top quark, the scalar top quarks, bottom quark, the scalar bottom 
quarks, tau lepton, the scalar tau leptons, $W$ boson, the charged Higgs boson, the charginos, $Z$ boson, the scalar 
Higgs bosons, and the neutralinos. 
We examine the effect of the contributions of the quartic terms and the pseudoscalar Higgs loops.
We find that the mass matrix of the neutral Higgs bosons in the MSSM at the one-loop level is evidently affected by the 
inclusion of the quartic terms and the pseudoscalar Higgs contribution. 

\section{The MSSM Higgs sector}

A starting point to evaluate the neutral Higgs boson masses in the MSSM with explicit CP violation may be the tree-level 
Higgs potential, which is given in terms of two Higgs doublets $H_i$ $(i = 1, 2)$ by
\begin{eqnarray}
	V^0 & = & {g_2^2\over 8} (H_1^{\dag} \vec\sigma H_1+ H_2^{\dag} \vec\sigma H_2)^2 + 
	{g_1^2\over 8}(|H_2|^2 - |H_1|^2)^2   \cr
	& &\mbox{} + m_1^2 |H_1|^2 + m_2^2 |H_2|^2 - m_3^2 (H_1^T \epsilon H_2 + {\rm H.c.}) \ ,
\end{eqnarray}
where $\epsilon$ is an antisymmetric $2 \times 2$  matrix with $\epsilon_{12} = 1$, 
$\vec\sigma$ denotes the three Pauli matrices, $g_1$ and $g_2$ are the U(1) and SU(2) gauge coupling constants, respectively, 
and $m_i^2$ $(i = 1, 2, 3)$ are the soft SUSY breaking masses.
Two soft SUSY breaking masses $m_i^2$ $(i = 1, 2)$ may be assumed to be real, without loss of generality.
We assume $m_3^2 = |m_3^2| e^{i \phi_3}$ to be complex.

Transforming to a unitary gauge, one may express two Higgs doublets in the mass eigenstates as
\begin{eqnarray}
\begin{array}{lll}
        H_1 & = & \displaystyle{{1 \over \sqrt{2}}} \left ( \begin{array}{c}
          v_1 +h_1 + i \sin \beta h_3   \cr
          \sin \beta C^{+ *}
  \end{array} \right )  \ ,  \cr
        H_2 & = & \displaystyle{{1 \over \sqrt{2}}} \left ( \begin{array}{c}
          \cos \beta C^+           \cr
          v_2 +h_2 + i \cos \beta h_3
  \end{array} \right ) e^{i \phi_0} \ ,  
\end{array}
\end{eqnarray}
where $h_1$, $h_2$, $h_3$ are three neutral Higgs fields, $C^+$ is the charged Higgs field, and $\phi_0$ is the relative phase $\phi_0$ between the two Higgs doublets.
In the presence of the explicit CP violation, $h_1$, $h_2$, $h_3$ are mixed states of the CP parity. 
With respect to these three neutral Higgs fields, we have the corresponding minimum conditions as  
\begin{eqnarray}
m_1^2 & = & m_3^2 \tan \beta \cos (\phi_3 + \phi_0) - {m_Z^2 \over 2} \cos 2 \beta \ , \cr
m_2^2 & = & m_3^2 \cot \beta \cos (\phi_3 + \phi_0) + {m_Z^2 \over 2} \cos 2 \beta \ , \cr
0 & = & m_3^2 \sin (\phi_3 + \phi_0) \ ,
\end{eqnarray}
respectively. 

At the tree level, the phases $\phi_3$ and $\phi_0$ can be set to zero.
Thus, it is well observed that there is no CP phase in the above tree-level Higgs potential. 
In order to accommodate the explicit CP violation scenario, the coefficients in the soft SUSY breaking terms 
are assumed to be complex in general. 
To be specific, we assume that the following quantities may be complex: 
The Higgs mixing parameters with mass dimension $\mu$, the U(1) gaugino mass $M_1$, the SU(2) gaugino mass $M_2$, 
and the trilinear soft SUSY breaking masses $A_t$, $A_b$, and $A_{\tau}$.
The complex phases in these quantities are responsible for the CP violation.
On the other hand, we take the soft SUSY breaking masses $m_Q$, $m_T$, $m_B$, $m_L$, and $m_E$ be real. 

Now, as the electroweak symmetry is broken spontaneously, the neutral Higgs fields develop non-trivial vacuum expectation values (VEVs).
The ratio of the two VEVs are defined as $\tan \beta = v_2/v_1$.
In terms of $v_1$ and $v_2$, the fermion masses are given as $m_t^2 = (h_t v_2)^2/2$ for top quark, 
$m_b^2  = (h_b v_1)^2/2$ for bottom quark, $m_{\tau}^2 = (h_{\tau} v_1)^2/2$ and the gauge boson masses 
as $m_W^2 = (g_2 v)^2/4$ and $m_Z^2 = (g^2_1 +g^2_2)v^2/4$ with $v^2 = v_1^2 + v_2^2$.

We calculate the radiative corrections to the tree-level Higgs sector.
We employ the effective potential method [9] to estimate radiative corrections to the tree-level Higgs sector.
The Higgs potential at the one-loop level is 
\[
        V = V^0 + V^1 \ ,
\]
where $V^1$ represents the radiative corrections due to various loops, and is calculated by the effective potential method. 
It may conveniently be decomposed as  
\begin{equation}
        V^1 = V^t +V^b +V^{\tau} +V^{\tilde \chi} +V^h +V^{{\tilde \chi}^0} \ ,
\end{equation}
where 
\[
V^t = \sum_{i = 1}^2 {3 {\cal M}_{{\tilde t}_i}^4 \over 32 \pi^2} 
  \left (\log {{\cal M}_{{\tilde t}_i}^2 \over \Lambda^2} - {3\over 2} \right )
  - {3 {\cal M}_t^4 \over 16 \pi^2} \left (\log {{\cal M}_t^2 \over \Lambda^2}
  - {3\over 2} \right ) \ , 
\]
for the one-loop contributions of top quark and the scalar top quarks, 
\[
V^b = \sum_{i = 1}^2 {3 {\cal M}_{{\tilde b}_i}^4 \over 32 \pi^2}
  \left (\log {{\cal M}_{{\tilde b}_i}^2 \over \Lambda^2} - {3\over 2} \right )
 -  {3 {\cal M}_b^4 \over 16 \pi^2} \left (\log {{\cal M}_b^2 \over \Lambda^2}
 - {3\over 2} \right ) \ , 
\]
for those of bottom quark and the scalar bottom quarks, 
\[
V^{\tau} = \sum_{i = 1}^2 {{\cal M}_{{\tilde \tau}_i}^4 \over 32 \pi^2}
  \left (\log {{\cal M}_{{\tilde \tau}_i}^2 \over \Lambda^2} - {3\over 2} \right )
     - {{\cal M}_{\tau}^4 \over 16 \pi^2} \left (\log {{\cal M}_{\tau}^2 \over \Lambda^2}
     - {3\over 2} \right ) \ , 
\]
for those of tau lepton and the scalar tau leptons,
\begin{eqnarray}
	V^{{\tilde \chi}} 
	& = &  {3 {\cal M}_W^4  \over 32 \pi^2} \left (\log {{\cal M}_W^2 \over \Lambda^2} - {3\over 2} \right ) 
        + {{\cal M}_{C^+}^4 \over 32 \pi^2} \left ( \log  {{\cal M}_{C^+}^2 \over \Lambda^2} - {3\over 2} \right )
        - \sum_{i = 1}^2 {{\cal M}_{{\tilde \chi}_i}^4 \over 16 \pi^2}
        \left (\log {{\cal M}_{{\tilde \chi}_i}^2 \over \Lambda^2} - {3\over 2} \right )   \ , \nonumber 
\end{eqnarray}
for those of $W$ boson, the charged Higgs boson and the charginos,
\[
V^h = {3 {\cal  M}_Z^4 \over  64 \pi^2}  
        \left (\log  {{\cal M}_Z^2 \over  \Lambda^2} - {3 \over 2}\right ) 
        + \sum_{i  = 1}^3 {{\cal  M}_{h_i}^4 \over 64 \pi^2} 
\left( \log {{\cal M}_{h_i}^2 \over \Lambda^2} - {3 \over 2} \right ) \ , 
\]
for those of $Z$ boson and both scalar and pseudoscalar neutral Higgs bosons,
and finally, 
\[
V^{{\tilde \chi}^0} = \mbox{} -
      \sum_{i = 1}^4 {{\cal M}_{{\tilde \chi}^0_i}^4 \over 32 \pi^2}
     \left (\log {{\cal M}_{{\tilde \chi}^0_i}^2 \over \Lambda^2} - {3\over 2} 
      \right ) \ ,
\]
for those of the neutralinos.

In the above expressions, $\Lambda$ is the renormalization scale in the modified minimal subtraction 
(${\overline {\rm MS}}$) scheme and ${\cal M}$ stands for the field-dependent mass matrix of the relevant ordinary 
particles and their superpartners.
Note that the coefficient of $V^{\tau}$ differs from those of  $V^t$ and $V^b$ as the color factor is absent in it. 
One may group the above components into two parts:  $V^t$, $V^b$, $V^{\tau}$, and $V^{\tilde \chi}$ into the charged 
part of $V^1$, whereas  $V^h$ and $V^{{\tilde \chi}^0}$ into the neutral part.  
Note that the loops of all the relevant particles and superpartners are included: top quark, bottom quark, tau lepton, 
$W$ boson, the charged Higgs boson, $Z$ boson, the neutral scalar and pseudoscalar Higgs bosons, and 
then the scalar top quarks, the scalar bottom quarks, the scalar tau leptons, the charginos, and the neutralinos.

In order to calculate the radiative corrections, one has to know the tree-level masses of particles and superpartners.
The scalar top quarks, the scalar bottom quarks, the scalar tau leptons, and the charginos obtain their tree-level 
masses as follows:
\begin{eqnarray}
m_{{\tilde t}_{1, 2}}^2 & = & m_t^2 + {1 \over 2}(m_Q^2 + m_T^2) + {m_Z^2 \over 4} \cos 2 \beta 
\mp \left [ \left \{ {1 \over 2} (m_Q^2 - m_T^2)   \right. \right. \cr
& &\mbox{}\left. \left. + \left ( {2 \over 3} m_W^2 - {5 \over 12} m_Z^2 \right ) \cos 2 \beta \right \}^2 
	+ m_t^2 (A_t^2  + \mu^2 \cot^2 \beta - 2 A_t \mu \cot \beta \cos \phi_t) \right ]^{1 \over 2}    \ , \cr
m_{{\tilde b}_{1, 2}}^2 & = & m_b^2 + {1 \over 2} (m_Q^2 + m_B^2) - {m_Z^2 \over 4} \cos 2 \beta 
\mp \left [ \left \{ {1 \over 2} (m_Q^2 - m_B^2) \right. \right. \cr
& &\mbox{}\left. \left. + \left ( {1 \over 12} m_Z^2 - {1 \over 3} m_W^2 \right ) \cos 2 \beta \right \}^2 
	+ m_b^2 (A_b^2 + \mu^2 \tan^2 \beta - 2 A_b \mu \tan \beta \cos \phi_b) \right ]^{1 \over 2}              \ , \cr
m_{{\tilde \tau}_{1, 2}}^2 & = & m_{\tau}^2 + {1 \over 2}(m_L^2 + m_E^2) - {m_Z^2 \over 4} \cos 2 \beta 
\mp \left [ \left \{ {1 \over 2} (m_L^2 - m_E^2) \right. \right. \cr
& &\mbox{}\left. \left. + \left ( {3 \over 8} m_Z^2 - {1 \over 2} m_W^2 \right ) \cos 2 \beta \right \}^2 
	+ m_{\tau}^2 (A_{\tau}^2 + \mu^2 \tan^2 \beta  
- 2 A_{\tau} \mu \tan \beta \cos \phi_{\tau}) \right ]^{1 \over 2}   \ , \cr
m_{{\tilde \chi}_{1, 2}}^2 & = & {1 \over 2} (M_2^2 + \mu^2) + m_W^2 
\mp \left [ \left \{ {1 \over 2} (M_2^2 - \mu^2) - m_W^2 \cos 2 \beta \right \}^2 \right. \cr
& &\left. \mbox{} + 2 m_W^2 \cos^2 \beta (M_2^2 + \mu^2 \tan^2 \beta + 
	2 M_2 \mu \tan \beta \cos \phi_c) \rule{0mm}{5.5mm} \right ]^{1\over 2}      \ ,
\end{eqnarray}
where $D$-terms are included.
Notice that four non-trivial CP phases appear in the above expressions. 
They are defined as follows: $\phi_t$ is the relative phase of $A_t$ and $\mu$, $\phi_b$ is that of $A_b$  and $\mu$, $\phi_{\tau}$ is that of $A_{\tau}$ and $\mu$, and $\phi_c$ is that of $M_2$ and $\mu$.

The tree-level neutralino mass matrix is given as
\begin{equation}
	{\cal M}_{{\tilde \chi}^0} = \left ( 
\displaystyle{
\begin{array}{cccc}
M_1 e^{i\phi_1} & 0 & -{g_1\over \sqrt2} H^0_1 & {g_1 \over \sqrt2} H^0_2 \cr
 & & & \cr
0 & M_2 & {g_2 \over \sqrt2} H^0_1 & - {g_2 \over \sqrt2} H^0_2   \cr  
 & & & \cr
- {g_1 \over \sqrt2} H^0_1 & {g_2\over \sqrt2} H^0_1 & 0 & -\mu e^{i\phi_2}  \cr  
 & & & \cr 
{g_1 \over \sqrt2} H^0_2 & - {g_2\over \sqrt2} H^0_2 & -\mu e^{i\phi_2} & 0 
\end{array}  } 
\right )  \ , 
\end{equation}
where additional complex phases appear: $\phi_1$ is the relative phase between $M_1$ and $M_2$ and 
$\phi_2$ is the relative phase between $M_2$ and $\mu$. However, $\phi_2$ is identical to $\phi_c$. 
Thus, the neutralino mass matrix introduces only one more phase.
Consequently, we have in general five non-trivial CP phases: 
$\phi_t$, $\phi_b$, $\phi_{\tau}$, $\phi_c = \phi_2$ and $\phi_1$.
The above neutralino mass matrix is complex and symmetric, but not Hermitian.
By diagonalizing the Hermitian matrix $M_{{\tilde \chi}^0}^{\dagger} M_{{\tilde \chi}^0}$ through a similarity transformation, the  tree-level neutralino masses are calculated. 
They are denoted as $m_{{\tilde \chi}_i^0}^2$ $(i = 1,2,3,4)$, and sorted 
such that $m^2_{{\tilde \chi}_i^0}<m^2_{{\tilde \chi}_j^0}$ for $i<j$.

For the Higgs sector, the tree-level mass of the pseudoscalar Higgs boson is obtained as 
\begin{equation}
m_{A^0}^2 = {2 m_3^2 \cos (\phi_3 + \phi_0) \over \sin 2 \beta}             \ .  
\end{equation}
Here we retain $\phi_3$ and $\phi_0$ because they become non-zero at the one loop level even though they are zero at the tree level.
The tree-level masses of the charged Higgs boson and the remaning neutral Higgs bosons are given as 
\begin{eqnarray}
m_{C^+}^2 & = & m_W^2 + m_{A^0}^2 \ , \cr 
& & \cr
m_{{h^0}, {H^0}}^2 & = & {1 \over 2} \left [m_Z^2 + m_{A^0}^2
    \mp \sqrt{ \left (m_Z^2 + m_{A^0}^2 \right )^2 - 4 m_Z^2 m_{A^0}^2 \cos^2 2 \beta}
 \right ]     \ , 
\end{eqnarray}
from the tree-level Higgs potential of the MSSM.
Within the context of perturbation theory, we eliminate both $m_1^2$ and $m_2^2$ in the one-loop functions of the Higgs 
bosons by using the tree-level minimum equations.

At the one loop level, there is one nontrivial mimimum condition with respect to $h_3$ field.
The CP-odd tadpole minimum equation is obtained as 
\begin{eqnarray}
0 & = & m_3^2 \sin (\phi_3 + \phi_0) + 
	{3 m_t^2 \mu A_t \sin \phi_t \over 16 \pi^2 v^2 \sin^2 \beta} f_1 (m_{{\tilde t}_1}^2,  \ m_{{\tilde t}_2}^2) 
+ {3 m_b^2 \mu A_b \sin \phi_b \over 16 \pi^2 v^2 \cos^2 \beta}
f_1 (m_{{\tilde b}_1}^2,  \ m_{{\tilde b}_2}^2) \cr
& &\mbox{} + {m_{\tau}^2 \mu A_{\tau} \sin \phi_{\tau} \over 16 \pi^2 v^2 \cos^2 \beta} 
	f_1 (m_{{\tilde \tau}_1}^2,  \ m_{{\tilde \tau}_2}^2) 
-  {m_W^2 \mu M_2 \sin \phi_c \over 4  \pi^2 v^2} f_1 (m_{{\tilde  \chi}_1}^2, \ m_{{\tilde \chi}_2}^2) \cr
& &\mbox{} + \sum^4_{k = 1} {m_{{\tilde \chi}^0_k}^2  \over 4 \pi^2 v^2} 
\left \{\log \left ({m_{{\tilde \chi}^0_k}^2 \over \Lambda^2} \right ) - 1 \right \} 
{E_3 \over \prod\limits_{a \not= k} (m_{{\tilde \chi}^0_k}^2 - m_{{\tilde \chi}^0_a}^2)}  \ , 
\end{eqnarray}
where
\begin{eqnarray}
	E_3 & = &\mbox{} - (m_{{\tilde \chi}^0_k}^2 - M_1^2) (m_{{\tilde \chi}^0_k}^2 - \mu^2 ) 
		M_2 \mu m_W^2 \sin \phi_2 \cr
		& &\mbox{} - (m_{{\tilde \chi}^0_k}^2 - M_2^2) (m_{{\tilde \chi}^0_k}^2 - \mu^2 ) 
		M_1 \mu (m_Z^2 - m_W^2) \sin(\phi_1 + \phi_2) \ . 
\end{eqnarray}
The dimensionless function $f_1$ is defined as
\[
	f_1 (m_x^2, \  m_y^2)  
        	= {2 \over (m_x^2-m_y^2)} \left \{m_x^2 \log  {m_x^2 \over 
        	\Lambda^2} -m_y^2 \log {m_y^2 \over \Lambda^2} \right \} - 2 \ .
\]
The six terms on the right-hand side of Eq. (9) come respectively from the tree-level Higgs potential, and the one-loop contributions of
the scalar top quark, the scalar bottom quark, the scalar tau lepton, the chargino, and the neutralinos.  

By differentiating the Higgs potential at the one-loop level with respect to the three neutral Higgs fields, a $3 \times 3$ symmetric 
mass matrix $M_{ij}$ is obtained in the $(h_1, h_2, h_3)$-basis. It may be decomposed as
\[
        M_{ij} = M^0_{ij} +M^1_{ij}
\]
where $M^0_{ij}$ is obtained from $V^0$, and $M^1_{ij}$ from $V^1$. 
In particular, $M^1_{ij}$ may further be decomposed as 
\[
        M^1_{ij} = M^t_{ij} + M^b_{ij} +M^{\tau}_{ij} +M_{ij}^{\tilde \chi} +M_{ij}^h +M_{ij}^{{\tilde \chi}^0}  
\]
where, as the superscripts suggest, $M^t_{ij}$ represents the radiative contributions from $V^t$. 
Likewise, $M^b_{ij}$, $M^{\tau}_{ij}$, $M^{\tilde \chi}_{ij}$, $M^h_{ij}$, and $M^{{\tilde \chi}^0}_{ij}$ represent 
respectively the radiative contributions from $V^b$, $V^{\tau}$, $V^{\tilde \chi}$, $V^h$, and $V^{{\tilde \chi}^0}$. 

The matrix elements of $M_{ij}$ in the $(h_1, h_2, h_3)$-basis are easily calculated as
\begin{eqnarray}
M_{11} & = & m_Z^2 \cos^2 \beta + {\bar m}_A^2 \sin^2 \beta + M_{11}^1 \ ,  \cr
M_{22} & = & m_Z^2 \sin^2 \beta + {\bar m}_A^2 \cos^2 \beta + M_{22}^1 \ ,  \cr
M_{33} & = & {\bar m}_A^2 + M_{33}^1 \ , \cr
M_{12} & = &\mbox{} - (m_Z^2 + {\bar m}_A^2) \cos \beta \sin \beta + M_{12}^1 \ ,  \cr
M_{13} & = & M_{13}^1 \ , \cr
M_{23} & = & M_{23}^1 \ . 
\end{eqnarray}
where the mass parameter ${\bar m}_A$ is defined as
\begin{eqnarray}
{\bar m}_A^2 & = & {2 \over \sin 2 \beta}
\left [\rule{0mm}{9mm} m_3^2 \cos (\phi_3 + \phi_0) 
+ {3 m_t^2 \mu A_t \cos \phi_t \over 16 \pi^2 v^2 \sin^2 \beta} f_1 (m_{{\tilde t}_1}^2,  \ m_{{\tilde t}_2}^2)  \right. \cr
& &\mbox{}\left. + {3 m_b^2 \mu A_b \cos \phi_b \over 16 \pi^2 v^2 \cos^2 \beta} f_1 (m_{{\tilde b}_1}^2,  \ m_{{\tilde b}_2}^2)
 + {m_{\tau}^2 \mu A_{\tau} \cos \phi_{\tau} \over 16 \pi^2 v^2 \cos^2 \beta} 
f_1 (m_{{\tilde \tau}_1}^2,  \ m_{{\tilde \tau}_2}^2)  \right. \cr
& &\mbox{}\left. +  {m_W^2 \mu M_2 \cos \phi_c \over 4  \pi^2 v^2} 
f_1 (m_{{\tilde  \chi}_1}^2, \ m_{{\tilde \chi}_2}^2) \right. \cr
& &\mbox{}\left. + \sum^4_{k = 1} {m_{{\tilde \chi}^0_k}^2  \over 4 \pi^2 v^2} 
\left \{\log \left ({m_{{\tilde \chi}^0_k}^2 \over \Lambda^2} \right ) - 1 \right \} 
{E_{33} \over \prod\limits_{a \not= k} (m_{{\tilde \chi}^0_k}^2 - m_{{\tilde \chi}^0_a}^2)}  \right ] \ ,
\end{eqnarray}
with 
\begin{eqnarray}
E_{33} & = & (m_{{\tilde \chi}^0_k}^2 - M_1^2) (m_{{\tilde \chi}^0_k}^2 - \mu^2 ) M_2 \mu m_W^2 \cos \phi_2 \cr
& &\mbox{} + (m_{{\tilde \chi}^0_k}^2 - M_2^2) (m_{{\tilde \chi}^0_k}^2 - \mu^2 ) 
M_1 \mu (m_Z^2 - m_W^2) \cos(\phi_1 + \phi_2) \ . 
\end{eqnarray}
In Eq. (11), the matrix elements $M_{i3}$ $(i = 1, 2)$ represent the mixing between the scalar and the pseudoscalar components.
Thus, $M_{i3} = M_{i3}^1$ $(i = 1, 2)$ indicate that the mixing occurs only at the one loop level. 
There is no mixing at the tree level. 

For $M^1_{ij}$, we first calculate $M_{ij}^h$ from $V^h$. 
The result is given as follows:
\begin{eqnarray}
M_{11}^h & = &\mbox{} -  {v^2 \cos^2 \beta \Delta_{h_1}^2 \over 32 \pi^2}
{f_2 (m_h^2, \ m_H^2) \over (m_H^2 - m_h^2)^2}  
- {m_Z^2 \over 32 \pi^2 v^2} (m_{A^0}^2 \sin^2 \beta - 4 m_Z^2 \cos^2 \beta) f_1 (m_h^2, \ m_H^2) \cr
& &\mbox{} + {m_Z^4 \cos^2 \beta \over 32 \pi^2 v^2} \log \left ({m_h^2  m_H^2  \over \Lambda^4} \right )  
+ {m_Z^2 \cos^2 \beta \Delta_{h_1} \over 16 \pi^2} {\log (m_H^2 / m_h^2) \over (m_H^2 - m_h^2)} \cr
& &\mbox{} + {m_Z^4 \cos^2 \beta \cos^2 2 \beta \over 32 \pi^2 v^2} 
	\log \left ({m_{A^0}^2 \over \Lambda^2} \right ) 
+ {m_Z^4 \cos^2 \beta \over 8 \pi^2 v^2} \log \left ({m_Z^6 \over \Lambda^6} \right ) \ , \cr
 & & \cr
M_{22}^h & = &\mbox{} - {v^2 \sin^2 \beta \Delta_{h_2}^2 \over 32 \pi^2}
{f_2 (m_h^2, \ m_H^2) \over (m_H^2 - m_h^2)^2}  
- {m_Z^2 \cos^2 \beta \over 32 \pi^2 v^2} (m_{A^0}^2 - 4 m_Z^2) f_1 (m_h^2, \ m_H^2) \cr
& &\mbox{} + {m_Z^4 \sin^2 \beta \over 32 \pi^2 v^2} \log \left ({m_h^2 m_H^2 \over \Lambda^4} \right )  
+ {m_Z^2 \sin^2  \beta \Delta_{h_2} \over 16 \pi^2}  {\log (m_H^2 /  m_h^2) \over (m_H^2  - m_h^2)} \cr
& &\mbox{} + {m_Z^4 \sin^2 \beta \cos^2 2 \beta \over 32 \pi^2 v^2} 
\log \left ({m_A^2 \over \Lambda^2} \right ) 
+ {m_Z^4 \sin^2 \beta \over 8 \pi^2 v^2} \log \left ({m_Z^6 \over \Lambda^6} \right ) \ , \cr
 & & \cr
M_{33}^h & = &\mbox{} - {m_Z^2 \over 32 \pi^2 v^2} 
\left [m_h^2 \left \{\log \left ({m_h^2 \over \Lambda^2} \right ) - 1 \right \} 
+ m_H^2 \left \{\log \left ({m_H^2 \over  \Lambda^2} \right ) - 1 \right \} \right ] \cr
& &\mbox{} - {m_Z^2 \over 64  \pi^2 v^2} \{m_Z^2 - m_{A^0}^2 (2 \sin^2 2 \beta - 3)\} f_1 (m_h^2, \ m_H^2) \cr
& &\mbox{} + {m_Z^2 m_{A^0}^2 \cos^2 2 \beta \over 16 \pi^2 v^2}
\left \{ \log \left ({m_A^2 \over \Lambda^2} \right ) - 1 \right \} \ , \cr
 & & \cr
M_{12}^h & = &\mbox{} - {v^2 \sin 2 \beta \Delta_{h_1} \Delta_{h_2} \over 64 \pi^2} 
{f_2 (m_h^2, \ m_H^2) \over (m_H^2 - m_h^2)^2}  
+ {m_Z^2 \sin 2 \beta \over 32 \pi^2 v^2} (m_{A^0}^2 -2 m_Z^2) f_1 (m_h^2, \ m_H^2) \cr
& &\mbox{} +{m_Z^4 \sin 2\beta \over 64 \pi^2 v^2} \log \left ({m_h^2 m_H^2  \over \Lambda^4} \right )  
+ {m_Z^2 \sin 2 \beta \over 64 \pi^2} (\Delta_{h_1} + \Delta_{h_2}) {\log (m_H^2 / m_h^2) \over (m_H^2 - m_h^2)} \cr
& &\mbox{} - {m_Z^4 \sin 2 \beta \cos^2 2 \beta \over 64 \pi^2 v^2}  
\log \left ({m_{A^0}^2 \over \Lambda^2} \right ) 
+ {m_Z^4 \sin 2 \beta \over 16 \pi^2 v^2} \log \left ({m_Z^6 \over \Lambda^6} \right ) \ ,  \cr
& & \cr
M_{13}^h & = & 0 \ , \cr
& & \cr
M_{23}^h & = & 0 \ ,
\end{eqnarray}
where
\begin{eqnarray}
&& f_2 (m_x^2,m_y^2) 
        = {m_x^2 + m_y^2 \over m_y^2 - m_x^2} \log {m_y^2 \over m_x^2} - 2 \ , \cr
&& \Delta_{h_1} = {2 m_Z^2 \over v^2} (m_Z^2 - m_{A^0}^2) \cos^2 \beta 
+ {4 m_Z^2 m_{A^0}^2 \over v^2} \sin^2 \beta  \ , \cr
&& \Delta_{h_2} =  {2 m_Z^2 \over v^2} (m_Z^2 - m_{A^0}^2) \sin^2 \beta 
+ {4 m_Z^2 m_{A^0}^2 \over v^2} \cos^2 \beta   \ .
\end{eqnarray}
In the above expressions, $m_{h^0}$, and $m_{H^0}$, and $m_{A^0}$ are the tree-level masses of the scalar and the 
pseudoscalar Higgs bosons.
Since there is no CP phase in $V^h$, we have $M_{13}^h = M_{23}^h = 0$ in Eq. (14).
Therefore, the explicit CP violation scenario in the Higgs sector of the MSSM may be regarded as the radiative CP violation, in the 
sense that the CP violation occurs only through the radiative corrections due to superpartners, via the CP phases in soft SUSY 
breaking terms.

Next, we calculate $M_{ij}^{{\tilde \chi}^0}$, the contributions of radiative corrections due to the four neutralinos. We obtain 
\begin{equation}
M_{ij}^{{\tilde \chi}^0}  = -\sum_{k=1}^4
{m_{{\tilde \chi}^0_k}^2 \over 16 \pi^2} 
\left (\log {m_{{\tilde \chi}^0_k}^2 \over \Lambda^2} - 1 \right ) 
{\partial^2 m_{{\tilde \chi}^0_k}^2 \over \partial h_i \partial h_j}
- \sum_{k=1}^4 {1\over 16  \pi^2} \log{m_{{\tilde \chi}^0_k}^2 \over \Lambda^2} \left 
({\partial m_{{\tilde \chi}^0_k}^2 \over \partial h_i}  \right)  
\left( {\partial  m_{{\tilde \chi}^0_k}^2 \over \partial h_j} \right )  \ , 
\end{equation}
where the first-order derivative $\partial m_{{\tilde \chi}^0_k}^2 / \partial h_i$ is given explicitly by
\[
        {\partial m_{{\tilde \chi}^0_k}^2 \over \partial h_i} 
        = -{\displaystyle 
        A_i m_{{\tilde \chi}^0_k}^6 +B_i m_{{\tilde \chi}^0_k}^4 
        +C_i m_{{\tilde \chi}^0_k}^2 +D_i 
        \over 
        \prod\limits^4_{a \not= k} (m_{{\tilde \chi}^0_k}^2 - m_{{\tilde \chi}^0_a}^2)}
  \\ ,
\]
and the second-order derivative $\partial^2 m_{{\tilde \chi}^0_k}^2 /\partial h_i \partial h_j$ by
\begin{eqnarray}
        {\partial^2 m_{{\tilde \chi}^0_k}^2 \over \partial h_i \partial h_j}  
        &=&\mbox{} -{A_{ij}m_{{\tilde \chi}^0_k}^6 +B_{ij} m_{{\tilde \chi}^0_k}^4 
        +C_{ij} m_{{\tilde \chi}^0_k}^2 +D_{ij} 
        \over \prod\limits^4_{a \not= k} 
        (m_{{\tilde \chi}^0_k}^2 - m_{{\tilde \chi}^0_a}^2)} \cr
& &\mbox{} 
        + \sum_{a\not=k}^4
{1 \over (m_{{\tilde \chi}^0_k}^2 - m_{{\tilde \chi}^0_a}^2)} 
        \left ({\partial m_{{\tilde \chi}^0_k}^2 \over \partial h_i} 
        {\partial m_{{\tilde \chi}^0_a}^2 \over \partial h_j} 
        +{\partial m_{{\tilde \chi}^0_a}^2 \over \partial h_i} 
        {\partial m_{{\tilde \chi}^0_k}^2 \over \partial h_j} \right ) \ .
\end{eqnarray}
The formulas for coefficients $A_i$, $B_i$, $C_i$, and $D_i$ $(i = 1, 2, 3)$ are given in Appendix A, 
and those for $A_{ij}$, $B_{ij}$, $C_{ij}$, and $D_{ij}$, $(i,j=1,2,3)$ are given in Appendix B. 
Actually, $A_{ij}$ = 0 for all $i,j$ because four elements of the neutralino mass matrix are zero.

We also carry out calculations for the remaining components of the one-loop mass matrices, namely, $M_{ij}^t$, $M_{ij}^b$, 
$M_{ij}^{\tau}$ and $M_{ij}^{\tilde \chi}$, and the results are given in Appendices C, D, E, and F, respectively. 

In the scenario of explicit CP violation, among the elements of the mass matrix for the neutral Higgs bosons, 
$M_{i3} = M^1_{i3} = M^t_{i3} + M^b_{i3} +M^{\tau}_{i3} +M_{i3}^{\tilde \chi} +M_{i3}^{{\tilde \chi}^0}$ $(i = 1, 2)$ 
are responsible for the scalar-pseudoscalar mixing and eventually account for the CP violation.
It is worthwhile to notice that our calculations show the general dependence of $M_{i3}$ on the various CP phases:
$M_{i3}^t$ $(i =1, 2)$ are proportional to $\sin \phi_t$, $M_{i3}^b$ $(i= 1,2)$ to $\sin \phi_b$, 
$M_{i3}^{\tau}$ $(i=1,2)$ to $\sin\phi_{\tau}$, $M_{i3}^{\tilde \chi}$ $(i = 1, 2)$ to $\sin \phi_c$.
Meanwhile, $M_{i3}^h$ $(i =1,2)$ are zero, as noticed above.
As for $M_{i3}^{{\tilde \chi}^0}$ $(i=1,2)$, they have complicated expressions, but one can easily see that every 
term is proportional to $\sin (\phi_1 +\phi_2)$ or $\sin \phi_1$ since $B_3$, $C_3$, $D_3$ are proportional to 
them whereas $B_{i3}$, $C_{i3}$, $D_{i3}$ $(i = 1, 2, 3)$ are zero.
Consequently, if any of the CP phases be not zero, they might give rise to the CP violation via the scalar-peudoscalar mixing
at the one-loop level.

\section{Numerical analysis}

We investigate the effects of CP phases in the one-loop corrected MSSM Higgs sector.
As we have emphasized, if there exist some CP phases in the Higgs potential at the one-loop level, they would produce 
the scalar-pseudoscalar mixing through non-zero $M_{13}$ and $M_{23}$ and thus generate explicit CP violation in 
the neutral Higgs sector.
In the MSSM Higgs sector, we show that there are in general five independent CP phases. 
They are contained in the masses of the scalar top quarks, the scalar bottom quarks, the scalar tau leptons, 
the charginos, and the neutralinos.

The mass matrix for the neutral Higgs bosons is derived by differentiating twice the Higgs potential with respect to the three Higgs fields. 
It is expressed as a 3 $\times$ 3 matrix. 
By calculating its three eigenvalues, one can obtain the three masses of the neutral Higgs bosons, analytically in principle.
However, we would not write down the full expressions for the neutral Higgs boson masses, since they are very complicated 
functions of many parameters coming from the soft SUSY breaking in the MSSM. 
Rather, we would like to analyze numerically the effect of the CP phases on the mass matrix of the neutral Higgs boson by focusing our attention to $M_{i3}$ $(i=1,2)$.
 
For the numerical calculations, we need concrete numbers for the parameter values.
We set $m_t = 175.0$ GeV, $m_b =4.0$ GeV, $m_{\tau} = 1.7$ GeV, $m_W = 80.4$ GeV, $m_Z = 91.1$ GeV, and 
$\sin^2 \theta_W$ = 0.231.
Then the remaining relevant free parameters are $\Lambda$, $\tan \beta$, $\mu$, $m_Q$, $m_T$, $m_B$, $m_L$, 
$m_E$, $A_t$, $A_b$, $A_{\tau}$, $M_1$, $M_2$, $\phi_t$, $\phi_b$, $\phi_{\tau}$, $\phi_c (\phi_2)$, and $\phi_1$.
Since for a moderate values of the $\tan \beta$, the contribution for the scalar bottom sector as well as the scalar tau 
lepton one is relatively small, we set $m_Q = m_L$, $m_T = m_B = m_E$, $A_t = A_b = A_{\tau}$.
At the electroweak scale one can take the relation of $M_1 = 5 \tan^2 \theta_W M_2/3$ between U(1) and SU(2) gaugino masses. 
Thus, we have $\Lambda$, $\tan \beta$, $m_Q$, $m_T$, $A_t$, $M_2$, and five CP phases as free parameters.
Note that we will choose different values for the soft SUSY breaking masses of SU(2) doublet and singlet scalar fermions
such that $m_Q =m_L$ is different from $m_T = m_B = m_E$ in order to consider large mass splitting between 
the left- and right- handed scalar fermions.

Now, as a typical illustration, we set $\phi_t = \phi_b = \phi_{\tau} = \phi_c (\phi_2) = \phi_1 = \pi/3$. 
We fix the renormalization scale $\Lambda$ in the effective potential at 300 GeV.
We set the remaning free parameters as ${\bar m}_A$ = 300 GeV, $m_Q$ = 800 GeV, $m_T = M_2$ = 400 GeV, and $A_t$ = 200 GeV.
With these input values, we calculate the elements of the neutral Higgs boson mass matrix for some values of
$\tan\beta$ and $\mu$. 

Our results are shown in Tables 1-8.
These Tables are divided into two categories: in Tables 1-4 the contributions from the quartic terms of $O(g_i^4)$ $(i= 1, 2)$ 
and the pseudoscalar Higgs contribution are excluded whereas they are included in Tables 5-8, in order to exhibit the effect of  
these  contributions.
Tables 1, 2, 5 and 6 displays numerical results for small $\tan \beta$ = 5 while Tables 3, 4, 7 and 8  for relatively large 
$\tan \beta$ = 30.
For Tables 1, 3, 5 and 7 we set $\mu$ = $- 400$ GeV while for Tables 2, 4, 6 and 8 we set $\mu$ = 400 GeV.
The entries of Tables are the elements of the mass matrix of the neutral Higgs bosons in the $(h_1, h_2, h_3)$-basis, 
where all the contributions are fully listed. The unit is (GeV)$^2$. 
Note that, as the CP phases induce the CP violation through the scalar-pseudoscalar mixing, the columns for 
$M_{13}$ and $M_{23}$ are filled with generally non-zero numbers.
The number in the last row in each column is the sum of the numbers in the preceding rows in the same column. 

In all of the Tables 1-8, one can easily notice that, among the contributions to $M_{13}$ and $M_{23}$, the tree-level ones and the Higgs ones vanish: $M^0_{13} = M^0_{23} =0$ and $M^h_{13} = M^h_{23} =0$. 
This reflects the fact that there is no CP phase in the tree-level Higgs sector of the MSSM.
All the other elements are not zero, accounting for the scalar-pseudoscalar mixing, because the CP phases are 
present there.  
Actually, one can trace the effects of the CP phases on various contributions to $M_{13}$ and $M_{23}$. 
If $\phi_t = 0$, one would have $M_{13}^t = M_{23}^t = 0$. 
Further, one would have $M_{13}^b = M_{23}^b = 0$ for $\phi_b = 0$.
Likewise, one would have $M_{13}^{\tau} = M_{23}^{\tau} = 0$ for $\phi_{\tau} = 0$, 
$M_{13}^{\tilde \chi} = M_{23}^{\tilde \chi} = 0$ for $\phi_c (\phi_2) = 0$, 
and $M_{13}^{{\tilde \chi}^0} = M_{23}^{{\tilde \chi}^0} = 0$ for $\phi_2 = \phi_1 = 0$.
Among the non-zero contributions to $M_{13}$ and $M_{23}$, we find that $M_{13}^{{\tilde \chi}^0}$ of Table 
with $\mu$ = $-400$ GeV contribute about 25 $\sim$ 30 \% to the scalar-pseudoscalar mixing. 
Thus, Tables 1, 3, 5, and 7 indicate that the radiative corrections due to the neutralino loops contribute roughly 
25 \% to the CP mixing between $h_1$ and $h_3$ components.

Now, let us  compare Table 1 with Table 5, Table 2 with Table 6, and so on to examine the effects of the contributions from the quartic terms of $O(g_i^4)$ $(i = 1, 2)$ and the contribution by the pseudoscalar Higgs boson $A^0$. 
It is quite evident that there are non-negligible differences between them. 
The effects can most clearly be seen in  $M_{33}^h$. 
In Tables 1-4,  we see that $M_{33}^h$ is identically zero. 
This is expected since neither $A^0$ nor quartic term contributions are included in Tables 1-4. 
On the other hand, Tables 5-8 show $M_{33}^h$ is definitely far from zero, indicating that both of them contribute in Tables 5-8. 
For the particular parameter values, $M_{33}^h$ is calculated as 55.9, 58.0, 30.2, and 39.2 (GeV)$^2$ in Tables 5,6,7, and 8, respectively.

The effects of the contributions from the quartic terms and $A^0$ are also seen in other entries, too.
By comparing Table 1 with Table 5, for example, one can see that those contributions alter the values of $M^h_{11}$, 
$M^h_{22}$, and $M^h_{12}$.
The value of $M_{33}^h$, as well as those of $M^h_{11}$, $M^h_{22}$, and $M^h_{12}$. is essential to evaluate the 
neutral Higgs boson masses from their mass matrix and the mixing among them, at the one-loop level. 
Consequently, these contributions should be included, as they play roles definitely in the matrix elements $M_{ij}^h$ at the one-loop level.
Without these contributions, the neutral Higgs boson masses at the one-loop level would be changed, if the changes be small. 
We further note that $M_{33}^h$ is in particular also affected by the inclusion of these contributions.

Now, to be concrete, we evaluate the neutral Higgs boson masses. 
From Table 5, we obtain $m_{h_i} (i = 1, 2, 3)$ = 109.8, 300.1, 302.6 GeV and we obtain  108.6, 300.1, 302.8 GeV from Table 6.
Since Tables 5 and 6 differ in the value of $\mu$, the dependence on $\mu$ is relatively insignificant, for $\tan \beta$ = 5.
For relatively large $\tan \beta$ = 30, the masses of the three neutral Higgs bosons are obtained as $m_{h_i} (i = 1, 2, 3)$ = 115.2, 300.0, 300.1 GeV for Table 7 ($\mu = -400$ GeV) and 115.0, 300.0, 300.2 GeV for Table 8 ($\mu$ = 400 GeV). 
We see that the mass of the lightest neutral Higgs boson becomes slightly larger for large $\tan\beta$.
All of Tables 5-8 produce the neutral Higgs boson masses without contradicting LEP2 data.  

We have some comments on the radiative corrections due to the loops of $Z$ boson, the neutral Higgs bosons, and the neutralinos.
One can observe in Table 5-8 that the sum of $M_{12}^h$ and $M_{12}^{{\tilde \chi}^0}$ are comparatively large. 
This implies that the contributions of $Z$ boson, the neutral Higgs bosons (both scalar and pseudoscalar), and neutralinos play important role in the radiative corrections to the mixing between $h_1$ and $h_2$ components at the one-loop level.
Especially, this mixing occurs most largely in Table 6 as $(M_{12}^h + M_{12}^{{\tilde \chi}^0}) = - 78.1$ $({\rm GeV})^2$.

The contributions of the neutralino loops depend crucially on the CP phase $\phi_1$.
In other words, the CP phase $\phi_1$ occurs only in the expressions for the neutralino contributions. 
Now, in order to examine in more detail the dependence of the contributions of the neutralino loops on $\phi_1$, we plot $M^{{\tilde \chi}^0}_{13}$, $M_{13}$, $M^{{\tilde \chi}^0}_{23}$, and $M_{23}$ as functions of $\phi_1$ in Fig. 1, 
where the values of other parameters are the same as Table 5.
We do not plot $M_{i3}^t$, $M_{i3}^b$ and $M_{i3}^{\tau}$ and $M_{i3}^{\tilde \chi}$ in Fig. 1, since their contributions are the same as those numerical values of Table 5.
In Table 5, the contributions of the loops of the scalar quark and the scalar tau lepton are almost null for $M_{13}$.
For $M_{13}$, the contributions of neutralino loops are larger than those of the scalar fermions for some values of $\phi_1$.
For both $M_{13}$ and $M_{23}$, Fig. 1 shows that the contributions of the neutralino loops are smaller than the chargino ones for the whole range of $0 < \phi_1 < 2 \pi$. 

\section{Conclusions}

In the MSSM, the tree-level neutral Higgs sector may be divided into the scalar part and pseudoscalar part, and 
there is no mixing between them. 
Any phase that can cause the scalar-pseudoscalar mixing, hence the CP violation, can be absorbed away at the tree level.
At the one-loop level, where explicit CP violation is viable by introducing several CP phases in the effective Higgs 
potential, the scalar-pseudoscalar mixing occurs in general.
The scalar-pseudoscalar mixing is manifested by the non-vanishing $M_{13}$ and $M_{23}$ matrix elements of the neutral Higgs bosons.
Evidently, these off-diagonal elements affect the masses of the neutral Higgs bosons when the mass matrix is diagonalized.

The mass matrix of the neutral Higgs boson in the MSSM is evaluated at the one-loop level with explicit CP violation. 
In explicit CP violation scenario, five non-trivial CP phases are introduced, from the soft SUSY breaking terms of the 
MSSM Lagrangian, in the masses of the scalar top quarks, the scalar bottom quarks, the scalar tau leptons, the charginos, 
and the neutralinos.
These phases penetrate into the mass matrix of the neutral Higgs bosons.
All the contributions of relevant loops are taken into account: 
The loops of the pseudoscalar Higgs boson as well as all the loops of top quark, the scalar top quarks, bottom quark, 
the scalar bottom quarks, tau lepton, the scalar tau leptons, $W$ boson, the charged Higgs boson, the charginos, 
$Z$ boson, the scalar Higgs bosons, and the neutralinos. 

Especially, we consider the contributions of the terms quartic in the electroweak coupling and the pseudoscalar Higgs loop 
contribution in the one-loop effective potential to the neutral Higgs boson masses at the MSSM with five explicit CP phases.
It is found that the contributions of quartic terms and $A^0$ are definitely non-zero to the (3,3)-element of the mass matrix of the neutral 
Higgs bosons.
The CP mixing between $h_1$ and $h_3$ components can be induced about 25 \% of the considered total contribution 
by the neutralino contributions.
The contributions of $Z$ boson, the neutral Higgs bosons (both scalar and pseudoscalar), and the neutralinos contribute 
largely to the mixing between $h_1$ and $h_2$ components above other ones.

\vskip 0.3 in

\noindent
{\large {\bf Acknowledgments}}
\vskip 0.2 in
\noindent
This work was supported by Korea Research Foundation Grant (2001-050-D00005).

\vskip 0.2 in

\vfil\eject
\noindent
{\large {\bf Appendix A}}
\vskip 0.2 in
\noindent
The coefficients that appear in the first derivatives of the neutralino masses with respect to the neutral Higgs fields in the radiatively corrected mass matrix for the neutral Higgs bosons are given as follows:
\begin{eqnarray}
A_1 & = &\mbox{} - {4 m_Z^2 \cos \beta \over v}   \ , \cr
B_1 & = &\mbox{} {4 M_1^2 m_W^2 \cos \beta \over v} 
      + {4 M_2^2 (m_Z^2 - m_W^2) \cos \beta \over v}
      + {4 M_Z^2 (m_Z^2 + \mu^2) \cos \beta \over v} \cr
    & &\mbox{} - {4 M_2 m_W^2 \mu \sin \beta \cos \phi_2 \over v} 
      + {4 M_1 \mu (m_W^2 - m_Z^2) \sin \beta \cos (\phi_1 + \phi_2) \over v} \ , \cr
C_1 & = &\mbox{}  - {4 m_Z^4 \mu^2 \sin \beta \sin 2 \beta \over v} 
      - {4 M_1^2 m_W^2 (m_W^2 + \mu^2) \cos \beta \over v} \cr
    & &\mbox{} + {4 M_2^2 (M_Z^2 - m_W^2) (m_W^2 - m_Z^2 - \mu^2) \cos \beta \over v} \cr
    & &\mbox{} + {8 M_1 M_2 m_W^2 (m_W^2 - m_Z^2) \cos \beta \cos \phi_1 \over v}     
      + {4 M_2 m_W^2 \mu (M_1^2 + \mu^2) \sin \beta \cos \phi_2 \over v}   \cr
    & &\mbox{} + {4 M_1 \mu (m_Z^2 - m_W^2) (M_2^2 + \mu^2) \sin \beta \cos (\phi_1 + \phi_2) \over v} 
      \ , \cr
D_1 & = &\mbox{} + {4 M_1^2 m_W^4 \mu^2 \sin \beta \sin 2 \beta \over v} 
      + {4 M_2^2 \mu^2 (m_Z^2 - m_W^2)^2 \sin \beta \sin 2 \beta \over v} \cr
    & &\mbox{} + {8 M_1 M_2 m_W^2 \mu^2 (m_Z^2 - m_W^2) \sin \beta \sin 2 \beta \cos \phi_1 \over v} 
      - {4 M_1^2 M_2 m_W^2 \mu^3 \sin \beta \cos \phi_2 \over v} \cr
    & &\mbox{} + {4 M_1 M_2^2 \mu^3 (m_W^2 - m_Z^2) \sin \beta \cos (\phi_1 + \phi_2) \over v}   \ ,
\end{eqnarray}
and
\begin{eqnarray}
A_2 &=& A_1 (\cos\beta \leftrightarrow \sin\beta)  \ , \cr
B_2 &=& B_1 (\cos\beta \leftrightarrow \sin\beta)  \ , \cr
C_2 &=& C_1 (\cos\beta \leftrightarrow \sin\beta)  \ , \cr
D_2 &=& D_1 (\cos\beta \leftrightarrow \sin\beta)  \ .
\end{eqnarray}
and
\begin{eqnarray}
A_3 &=& 0   \ , \cr
B_3 &=&\mbox{} - {4 M_2 m_W^2 \mu \sin \phi_2 \over v} 
    + {4 M_1 \mu (m_W^2 - m_Z^2) \sin (\phi_1 + \phi_2) \over v} \ , \cr
C_3 &=& {4 M_2 m_W^2 \mu (M_1^2 + \mu^2) \sin \phi_2 \over v} 
    + {4 M_1 \mu (m_Z^2 - m_W^2) (M_2^2 +\mu^2) \sin (\phi_1 + \phi_2) \over v}  \ , \cr
D_3 &=&\mbox{} - {4 M_1^2 M_2 m_W^2 \mu^3 \sin \phi_2 \over v} 
    + {4 M_1 M_2^2 \mu^3 (m_W^2 - m_Z^2) \sin (\phi_1 + \phi_2) \over v} \ .
\end{eqnarray}
\vskip 0.2 in
\noindent
{\large {\bf Appendix B}}
\vskip 0.2 in
\noindent
The coefficients that appear in the second derivatives of the neutralino masses with respect to the neutral Higgs fields in the radiatively corrected mass matrix for the neutral Higgs bosons are given as follows:
\begin{eqnarray}
A_{11} & = & 0  \ , \cr
B_{11} & = & {8 m_Z^4 \cos^2 \beta \over v^2}   \ , \cr
C_{11} & = &\mbox{} - {8  M_1^2 m_W^4  \cos^2 \beta \over  v^2} 
-  {8 M_2^2 (m_Z^2 - m_W^2)^2 \cos^2 \beta \over v^2} \cr
& &\mbox{} + {16 M_1 M_2 m_W^2 (m_W^2 - m_Z^2) \cos^2 \beta \cos \phi_1 \over v^2}  \ , \cr
D_{11} & = & 0  \  , 
\end{eqnarray}
and
\begin{eqnarray}
A_{22} & = & A_{11} (\cos\beta \leftrightarrow \sin\beta) = 0  \ , \cr
B_{22} & = & B_{11} (\cos\beta \leftrightarrow \sin\beta)  \ , \cr
C_{22} & = & C_{11} (\cos\beta \leftrightarrow \sin\beta)  \ , \cr
D_{22} & = & D_{11} (\cos\beta \leftrightarrow \sin\beta) = 0  \ , 
\end{eqnarray}
and
\begin{eqnarray}
A_{12} & = & 0  \ , \cr
B_{12} & = & {4 m_Z^4 \sin 2 \beta \over v^2} \ , \cr
C_{12} & = &\mbox{} - {4 M_1^2 m_W^4 \sin 2 \beta \over v^2}
 - {8 m_Z^4 \mu^2 \sin 2 \beta \over v^2} 
- {4 M_2^2 (m_Z^2 - m_W^2)^2 \sin 2 \beta \over v^2}  \cr
& &\mbox{} + {8 M_1 M_2 m_W^2 (m_W^2  - m_Z^2) \sin 2 \beta \cos \phi_1 \over  v^2}  \ , \cr
D_{12} & = &{8 M_1^2 m_W^4 \mu^2 \sin 2  \beta \over v^2} + 
{8 M_2^2  \mu^2 (m_Z^2 - m_W^2)^2 \sin 2 \beta \over v^2} \cr 
& &\mbox{} + {16 M_1 M_2 m_W^2 \mu^2 (m_Z^2 - m_W^2) \sin 2 \beta \cos \phi_1 \over v^2}  \ , 
\end{eqnarray}
and $A_{l3} = B_{l3} = C_{l3} = D_{l3} = 0$  for $l = 1, 2, 3$.
\vskip 0.2 in
\noindent
{\large {\bf Appendix C}}
\vskip 0.2 in
\noindent
The elements for the mass matrix of the neutral Higgs bosons due to the radiative contributions of the top quark and 
scalar top quarks are   
\begin{eqnarray}
M_{11}^t & = &\mbox{} - {3 \over 4 \pi^2 v^2}
\left \{ {m_t^2 \mu \Delta_{{\tilde t}_1} \over \sin \beta}
+ {\cos \beta \Delta_{\tilde t} \over 2} \right \}^2
{f_2 (m_{{\tilde t}_1}^2, \ m_{{\tilde t}_2}^2) \over
(m_{{\tilde t}_2}^2 - m_{{\tilde t}_1}^2)^2}
+ {3 m_Z^4 \cos^2 \beta \over 64 \pi^2 v^2}
\log \left ({m_{{\tilde t}_1}^2  m_{{\tilde t}_2}^2 \over \Lambda^4} \right ) \cr
& & \cr
& &\mbox{} + {3 \cos^2 \beta \over 16 \pi^2 v^2}
\left( {4 m_W^2 \over 3} - {5 m_Z^2 \over 6} \right)^2 
f_1 (m_{{\tilde t}_1}^2, \ m_{{\tilde t}_2}^2)   \cr
& & \cr
& &\mbox{} + {3 m_Z^2 \cos \beta \over 8 \pi^2 v^2}
\left \{ {m_t^2 \mu \Delta_{{\tilde t}_1} \over \sin \beta}
+ {\cos \beta \Delta_{\tilde t} \over 2} \right \}
{\log (m_{{\tilde t}_2}^2 / m_{{\tilde t}_1}^2)
 \over (m_{{\tilde t}_2}^2 - m_{{\tilde t}_1}^2)}   \cr
& & \cr
M_{22}^t & = &\mbox{} - {3 \over 4 \pi^2 v^2}
\left \{ {m_t^2 A_t \Delta_{{\tilde t}_2} \over \sin \beta}
- {\sin \beta \Delta_{\tilde t} \over 2} \right \}^2
{f_2 (m_{{\tilde t}_1}^2, \ m_{{\tilde t}_2}^2) \over
(m_{{\tilde t}_2}^2 - m_{{\tilde t}_1}^2)^2}
- {3 m_t^4 \over 2 \pi^2 v^2 \sin^2 \beta} 
\log \left ({m_t^2 \over \Lambda^2} \right ) \cr
& & \cr
& &\mbox{} + {3 \sin^2 \beta \over 16 \pi^2 v^2}
\left( {4 m_W^2 \over 3} - {5 m_Z^2 \over 6} \right)^2 
f_1 (m_{{\tilde t}_1}^2, \ m_{{\tilde t}_2}^2)   \cr
& & \cr
& &\mbox{} + {3 \sin \beta \over 8 \pi^2 v^2}
\left ({4 m_t^2 \over \sin^2 \beta} - m_Z^2 \right)
\left \{ {m_t^2 A_t \Delta_{{\tilde t}_2} \over \sin \beta} 
- {\sin \beta \Delta_{\tilde t} \over 2} \right \}
{\log (m_{{\tilde t}_2}^2 / m_{{\tilde t}_1}^2)
 \over (m_{{\tilde t}_2}^2 - m_{{\tilde t}_1}^2)} \cr
& & \cr
& &\mbox{} + {3 \over 16 \pi^2 v^2}
\left ({2 m_t^2 \over \sin \beta} - {\sin \beta m_Z^2 \over 2} \right)^2
\log \left ({m_{{\tilde t}_1}^2  m_{{\tilde t}_2}^2 \over \Lambda^4} \right ) \cr
& & \cr
M_{33}^t & = &\mbox{} - {3 m_t^4 \mu^2 A_t^2 \sin^2 \phi_t \over 4 \pi^2 v^2 \sin^4 \beta}
{f_2 (m_{{\tilde t}_1}^2, \ m_{{\tilde t}_2}^2) \over (m_{{\tilde t}_2}^2 - m_{{\tilde t}_1}^2 )^2} \ ,  \cr
& & \cr
M_{12}^t & = &\mbox{} - {3 \over 4 \pi^2 v^2}
\left \{ {m_t^2 \mu \Delta_{{\tilde t}_1} \over \sin \beta}
+ {\cos \beta \Delta_{\tilde t} \over 2} \right \}
\left \{ {m_t^2 A_t \Delta_{{\tilde t}_2} \over \sin \beta}
- {\sin \beta \Delta_{\tilde t} \over 2} \right \}
{f_2 (m_{{\tilde t}_1}^2, \ m_{{\tilde t}_2}^2)
\over (m_{{\tilde t}_2}^2 - m_{{\tilde t}_1}^2)^2}  \cr
& & \cr
& & \mbox{} - {3 \sin 2 \beta \over 32 \pi^2 v^2}
\left ({4 m_W^2 \over 3} - {5 m_Z^2 \over 6} \right)^2 
f_1 (m_{{\tilde t}_1}^2, \ m_{{\tilde t}_2}^2)  \cr
& & \cr
& &\mbox{} + {3 \sin \beta \over 16 \pi^2 v^2}
\left ({4 m_t^2 \over \sin^2 \beta} - m_Z^2 \right)
\left \{ {m_t^2 \mu \Delta_{{\tilde t}_1} \over \sin \beta}
+ {\cos \beta \Delta_{\tilde t} \over 2} \right \}
{\log (m_{{\tilde t}_2}^2 / m_{{\tilde t}_1}^2)
 \over (m_{{\tilde t}_2}^2 - m_{{\tilde t}_1}^2)} \cr
& & \cr
& &\mbox{} + {3 m_Z^2 \cos \beta \over 16 \pi^2 v^2} 
\left \{ {m_t^2 A_t \Delta_{{\tilde t}_2} \over \sin \beta}
- {\sin \beta \Delta_{\tilde t} \over 2} \right \}
{\log (m_{{\tilde t}_2}^2 / m_{{\tilde t}_1}^2)
 \over (m_{{\tilde t}_2}^2 - m_{{\tilde t}_1}^2)} \cr
& & \cr
& &\mbox{} + {3 m_Z^2 \sin 2 \beta \over 128 \pi^2 v^2}
\left ({4 m_t^2 \over \sin^2 \beta} - m_Z^2 \right)
\log \left ({m_{{\tilde t}_1}^2 m_{{\tilde t}_2}^2 \over \Lambda^4} \right ) \cr
& & \cr
M_{13}^t & = & \mbox{} - {3 m_t^2 \mu A_t \sin \phi_t \over 4 \pi^2 v^2 \sin^2 \beta}
\left \{{m_t^2 \mu \Delta_{{\tilde t}_1} \over \sin \beta}
+ {\cos \beta \Delta_{\tilde t} \over 2} \right \}
{f_2 (m_{{\tilde t}_1}^2, \ m_{{\tilde t}_2}^2) \over (m_{{\tilde t}_2}^2 - m_{{\tilde t}_1}^2)^2 }  \cr
& & \cr
& &\mbox{}
+ {3 m_t^2 m_Z^2 \mu A_t \cot \beta \sin \phi_t \over 16 \pi^2 v^2 \sin \beta}
{\log (m_{{\tilde t}_2}^2 / m_{{\tilde t}_1}^2) \over (m_{{\tilde t}_2}^2 - m_{{\tilde t}_1}^2)} \ , \cr
& & \cr
M_{23}^t & = & \mbox{} - {3 m_t^2 \mu A_t \sin \phi_t
\over 4 \pi^2 v^2 \sin^2 \beta}
\left \{{m_t^2 A_t \Delta_{{\tilde t}_2} \over \sin \beta}
- {\sin \beta \Delta_{\tilde t} \over 2} \right \}
{f_2 (m_{{\tilde t}_1}^2, \ m_{{\tilde t}_2}^2) \over
(m_{{\tilde t}_2}^2 - m_{{\tilde t}_1}^2)^2 }  \cr
& & \cr
& &\mbox{}
+ {3 m_t^2 \mu A_t \sin \phi_t \over 16 \pi^2 v^2 \sin \beta}
\left ( {4 m_t^2 \over \sin^2 \beta} - m_Z^2 \right )
{ \log (m_{{\tilde t}_2}^2 / m_{{\tilde t}_1}^2) \over
(m_{{\tilde t}_2}^2 - m_{{\tilde t}_1}^2)}    \ ,
\end{eqnarray}
with
\begin{eqnarray}
\Delta_{{\tilde t}_1} & = & \mu \cot \beta - A_t \cos \phi_t \ , \cr
\Delta_{{\tilde t}_2} & = & A_t - \mu \cot \beta \cos \phi_t \ , \cr
\Delta_{\tilde t} & = & \left ({4 \over 3} m_W^2 - {5 \over 6} m_Z^2 \right) 
\left \{ m_Q^2 - m_T^2 + \left ({4 \over 3} m_W^2 - {5 \over 6} m_Z^2 \right) \cos 2 \beta  \right \}  \ .
\end{eqnarray}
\vskip 0.2 in
\noindent
{\large {\bf Appendix D}}
\vskip 0.2 in
\noindent
The elements for the mass matrix of the neutral Higgs bosons due to the radiative contributions of the bottom quark and 
scalar bottom quarks are 
\begin{eqnarray}
M_{11}^b & = &\mbox{} - {3 \over 4 \pi^2 v^2}
\left \{ {m_b^2 A_b \Delta_{{\tilde b}_1} \over \cos \beta}
+ {\cos \beta \Delta_{\tilde b} \over 2} \right \}^2
{f_2 (m_{{\tilde b}_1}^2, \ m_{{\tilde b}_2}^2) \over
(m_{{\tilde b}_2}^2 - m_{{\tilde b}_1}^2)^2}
- {3 m_b^4 \over 2 \pi^2 v^2 \cos^2 \beta} 
\log \left ({m_b^2 \over \Lambda^2} \right ) \cr
& & \cr
& &\mbox{} + {3 \cos^2 \beta \over 16 \pi^2 v^2}
\left( {m_Z^2 \over 6} - {2 m_W^2 \over 3} \right)^2 
f_1 (m_{{\tilde b}_1}^2, \ m_{{\tilde b}_2}^2)   \cr
& & \cr
& &\mbox{} + {3 \cos \beta \over 8 \pi^2 v^2}
\left ({4 m_b^2 \over \cos^2 \beta} - m_Z^2 \right)
\left \{ {m_b^2 A_b \Delta_{{\tilde b}_1} \over \cos \beta} 
+ {\cos \beta \Delta_{\tilde b} \over 2} \right \}
{\log (m_{{\tilde b}_2}^2 / m_{{\tilde b}_1}^2)
 \over (m_{{\tilde b}_2}^2 - m_{{\tilde b}_1}^2)} \cr
& & \cr
& &\mbox{} + {3 \over 16 \pi^2 v^2}
\left ({2 m_b^2 \over \cos \beta} - {m_Z^2 \cos \beta \over 2} \right)^2
\log \left ({m_{{\tilde b}_1}^2  m_{{\tilde b}_2}^2 \over \Lambda^4} \right ) \ , \cr
& & \cr
M_{22}^b & = &\mbox{} - {3 \over 4 \pi^2 v^2}
\left \{ {m_b^2 \mu \Delta_{{\tilde b}_2} \over \cos \beta}
- {\sin \beta \Delta_{\tilde b} \over 2} \right \}^2
{f_2 (m_{{\tilde b}_1}^2, \ m_{{\tilde b}_2}^2) \over (m_{{\tilde b}_2}^2 - m_{{\tilde b}_1}^2)^2}
+ {3 m_Z^4 \sin^2 \beta \over 64 \pi^2 v^2}
\log \left ({m_{{\tilde b}_1}^2  m_{{\tilde b}_2}^2 \over \Lambda^4} \right ) \cr
& & \cr
& &\mbox{}+ {3 \sin^2 \beta \over 16 \pi^2 v^2}
\left( {m_Z^2 \over 6} - {2 m_W^2 \over 3} \right)^2 
f_1 (m_{{\tilde b}_1}^2, \ m_{{\tilde b}_2}^2)   \cr
& & \cr
& &\mbox{} + {3 m_Z^2 \sin \beta \over 8 \pi^2 v^2}
\left \{ {m_b^2 \mu \Delta_{{\tilde b}_2} \over \cos \beta}
- {\sin \beta \Delta_{\tilde b} \over 2} \right \}
{\log (m_{{\tilde b}_2}^2 / m_{{\tilde b}_1}^2)
 \over (m_{{\tilde b}_2}^2 - m_{{\tilde b}_1}^2)}   \ , \cr
& & \cr 
M_{33}^b & = &\mbox{} - {3 m_b^4 \mu^2 A_b^2 \sin^2 \phi_b \over 4 \pi^2 v^2 \cos^4 \beta}
{f_2 (m_{{\tilde b}_1}^2, \ m_{{\tilde b}_2}^2) \over (m_{{\tilde b}_2}^2
- m_{{\tilde b}_1}^2 )^2}  \ , \cr
& & \cr
M_{12}^b & = &\mbox{} - {3 \over 4 \pi^2 v^2}
\left \{ {m_b^2 A_b \Delta_{{\tilde b}_1} \over \cos \beta}
+ {\cos \beta \Delta_{\tilde b} \over 2} \right \}
\left \{ {m_b^2 \mu \Delta_{{\tilde b}_2} \over \cos \beta}
- {\sin \beta \Delta_{\tilde b} \over 2} \right \}
{f_2 (m_{{\tilde b}_1}^2, \ m_{{\tilde b}_2}^2)
\over (m_{{\tilde b}_2}^2 - m_{{\tilde b}_1}^2)^2}  \cr
& & \cr
& & \mbox{} - {3 \sin 2 \beta \over 32 \pi^2 v^2}
\left ({m_Z^2 \over 6} - {2 m_W^2 \over 3} \right)^2 
f_1 (m_{{\tilde b}_1}^2, \ m_{{\tilde b}_2}^2)  \cr
& & \cr
& &\mbox{} + {3 \cos \beta \over 16 \pi^2 v^2}
\left ({4 m_b^2 \over \cos^2 \beta} - m_Z^2 \right)
\left \{ {m_b^2 \mu \Delta_{{\tilde b}_2} \over \cos \beta}
- {\sin \beta \Delta_{\tilde b} \over 2} \right \}
{\log (m_{{\tilde b}_2}^2 / m_{{\tilde b}_1}^2)
 \over (m_{{\tilde b}_2}^2 - m_{{\tilde b}_1}^2)} \cr
& & \cr
& &\mbox{} + {3 m_Z^2 \sin 2 \beta \over 32 \pi^2 v^2} 
\left \{ {m_b^2 A_b \Delta_{{\tilde b}_1} \over \cos^2 \beta}
+ {\Delta_{\tilde b} \over 2} \right \}
{\log (m_{{\tilde b}_2}^2 / m_{{\tilde b}_1}^2)
 \over (m_{{\tilde b}_2}^2 - m_{{\tilde b}_1}^2)} \cr
& & \cr
& &\mbox{} + {3 m_Z^2 \sin 2 \beta \over 128 \pi^2 v^2}
\left ({4 m_b^2 \over \cos^2 \beta} - m_Z^2 \right)
\log \left ({m_{{\tilde b}_1}^2 m_{{\tilde b}_2}^2 \over \Lambda^4} \right ) \ , \cr
& & \cr
M_{13}^b & = & \mbox{} - {3 m_b^2 \mu A_b \sin \phi_b \over 4 \pi^2 v^2 \cos^2 \beta}
\left \{{m_b^2 A_b \Delta_{{\tilde b}_1} \over \cos \beta}
+ {\cos \beta \Delta_{\tilde b} \over 2} \right \}
{f_2 (m_{{\tilde b}_1}^2, \ m_{{\tilde b}_2}^2) \over
(m_{{\tilde b}_2}^2 - m_{{\tilde b}_1}^2)^2 }  \cr
& & \cr
& &\mbox{} + {3 m_b^2 \mu A_b \sin \phi_b \over 16 \pi^2 v^2 \cos \beta}
\left ( {4 m_b^2 \over \cos^2 \beta} - m_Z^2 \right )
{ \log (m_{{\tilde b}_2}^2 / m_{{\tilde b}_1}^2) \over (m_{{\tilde b}_2}^2 - m_{{\tilde b}_1}^2)} \ , \cr
& & \cr
M_{23}^b & = &\mbox{} - {3 m_b^2 \mu A_b \sin \phi_b \over 4 \pi^2 v^2 \cos^2 \beta}
\left \{{m_b^2 \mu \Delta_{{\tilde b}_2} \over \cos \beta}
- {\sin \beta \Delta_{\tilde b} \over 2} \right \}
{f_2 (m_{{\tilde b}_1}^2, \ m_{{\tilde b}_2}^2) \over
(m_{{\tilde b}_2}^2 - m_{{\tilde b}_1}^2)^2 }  \cr
& & \cr
& &\mbox{}
+ {3 m_b^2 m_Z^2 \mu A_b \tan \beta \sin \phi_b \over 16 \pi^2 v^2 \cos \beta}
{\log (m_{{\tilde b}_2}^2 / m_{{\tilde b}_1}^2)
\over (m_{{\tilde b}_2}^2 - m_{{\tilde b}_1}^2)}  \ , 
\end{eqnarray}
with
\begin{eqnarray}
\Delta_{{\tilde b}_1} & = & A_b - \mu \tan \beta \cos \phi_b \ , \cr
\Delta_{{\tilde b}_2} & = & \mu \tan \beta - A_b \cos \phi_b  \ , \cr
\Delta_{\tilde b} & = & \left ({1 \over 6} m_Z^2 - {2 \over 3} m_W^2 \right) 
\left \{ m_Q^2 - m_B^2 + \left ({1 \over 6} m_Z^2 - {2 \over 3} m_W^2 \right) \cos 2 \beta  \right \}  \ .
\end{eqnarray}
\vskip 0.2 in
\noindent
{\large {\bf Appendix E}}
\vskip 0.2 in
\noindent
The elements for the mass matrix of the neutral Higgs bosons due to the radiative contributions of the tau lepton and 
scalar tau leptons are
\begin{eqnarray}
M_{11}^{\tau} & = &\mbox{} - {1 \over 4 \pi^2 v^2}
\left \{ {m_{\tau}^2 A_{\tau} \Delta_{{\tilde \tau}_1} \over \cos \beta}
+ {\cos \beta \Delta_{\tilde \tau} \over 2} \right \}^2
{f_2 (m_{{\tilde \tau}_1}^2, \ m_{{\tilde \tau}_2}^2) \over
(m_{{\tilde \tau}_2}^2 - m_{{ \tilde \tau}_1}^2)^2}
- {m_{\tau}^4 \over 2 \pi^2 v^2 \cos^2 \beta} 
\log \left ({m_{\tau}^2 \over \Lambda^2} \right ) \cr
& & \cr
& &\mbox{}+ {\cos^2 \beta \over 16 \pi^2 v^2}
\left( {3 m_Z^2 \over 4} - m_W^2 \right)^2 
f_1 (m_{{\tilde \tau}_1}^2, \ m_{{\tilde \tau}_2}^2)   \cr
& & \cr
& &\mbox{} + {\cos \beta \over 8 \pi^2 v^2}
\left ({4 m_{\tau}^2 \over \cos^2 \beta} - m_Z^2 \right)
\left \{ {m_{\tau}^2 A_{\tau} \Delta_{{\tilde \tau}_1} \over \cos \beta} 
+ {\cos \beta \Delta_{\tilde \tau} \over 2} \right \}
{\log (m_{{\tilde \tau}_2}^2 / m_{{\tilde \tau}_1}^2)
 \over (m_{{\tilde \tau}_2}^2 - m_{{\tilde \tau}_1}^2)} \cr
& & \cr
& &\mbox{} + {1 \over 16 \pi^2 v^2}
\left ({2 m_{\tau}^2 \over \cos \beta} - {m_Z^2 \cos \beta \over 2} \right)^2
\log \left ({m_{{\tilde \tau}_1}^2  m_{{\tilde \tau}_2}^2 \over \Lambda^4} \right ) \ , \cr
& & \cr
M_{22}^{\tau} & = &\mbox{} - {1 \over 4 \pi^2 v^2}
\left \{ {m_{\tau}^2 \mu \Delta_{{\tilde \tau}_2} \over \cos \beta}
- {\sin \beta \Delta_{\tilde \tau} \over 2} \right \}^2
{f_2 (m_{{\tilde \tau}_1}^2, \ m_{{\tilde \tau}_2}^2) \over
(m_{{\tilde \tau}_2}^2 - m_{{\tilde \tau}_1}^2)^2}
+ {m_Z^4 \sin^2 \beta \over 64 \pi^2 v^2}
\log \left ({m_{{\tilde \tau}_1}^2  m_{{\tilde \tau}_2}^2 \over \Lambda^4} \right ) \cr
& & \cr
& &\mbox{}+ {\sin^2 \beta \over 16 \pi^2 v^2}
\left( {3 m_Z^2 \over 4} - m_W^2 \right)^2 
f_1 (m_{{\tilde \tau}_1}^2, \ m_{{\tilde \tau}_2}^2)   \cr
& & \cr
& &\mbox{} + {m_Z^2 \sin \beta \over 8 \pi^2 v^2}
\left \{ {m_{\tau}^2 \mu \Delta_{{\tilde \tau}_2} \over \cos \beta}
- {\sin \beta \Delta_{\tilde \tau} \over 2} \right \}
{\log (m_{{\tilde \tau}_2}^2 / m_{{\tilde \tau}_1}^2)
 \over (m_{{\tilde \tau}_2}^2 - m_{{\tilde \tau}_1}^2)}   \ , \cr
& & \cr 
M_{33}^{\tau} & = &\mbox{} - {m_{\tau}^4 \mu^2 A_{\tau}^2 \sin^2 \phi_{\tau} \over 4 \pi^2 v^2 \cos^4 \beta}
{f_2 (m_{{\tilde \tau}_1}^2, \ m_{{\tilde \tau}_2}^2) \over (m_{{\tilde \tau}_2}^2 - m_{{\tilde \tau}_1}^2 )^2} \ , \cr
& & \cr
M_{12}^{\tau} & = &\mbox{} - {1 \over 4 \pi^2 v^2}
\left \{ {m_{\tau}^2 A_{\tau} \Delta_{{\tilde \tau}_1} \over \cos \beta}
+ {\cos \beta \Delta_{\tilde \tau} \over 2} \right \}
\left \{ {m_{\tau}^2 \mu \Delta_{{\tilde \tau}_2} \over \cos \beta}
- {\sin \beta \Delta_{\tilde \tau} \over 2} \right \}
{f_2 (m_{{\tilde \tau}_1}^2, \ m_{{\tilde \tau}_2}^2)
\over (m_{{\tilde \tau}_2}^2 - m_{{\tilde \tau}_1}^2)^2}  \cr
& & \cr
& & \mbox{} - {\sin 2 \beta \over 32 \pi^2 v^2}
\left ({3 m_Z^2 \over 4} - m_W^2 \right)^2 
f_1 (m_{{\tilde \tau}_1}^2, \ m_{{\tilde \tau}_2}^2)  \cr
& & \cr
& &\mbox{} + {\cos \beta \over 16 \pi^2 v^2}
\left ({4 m_{\tau}^2 \over \cos^2 \beta} - m_Z^2 \right)
\left \{ {m_{\tau}^2 \mu \Delta_{{\tilde \tau}_2} \over \cos \beta}
- {\sin \beta \Delta_{\tilde \tau} \over 2} \right \}
{\log (m_{{\tilde \tau}_2}^2 / m_{{\tilde \tau}_1}^2)
 \over (m_{{\tilde \tau}_2}^2 - m_{{\tilde \tau}_1}^2)} \cr
& & \cr
& &\mbox{} + {m_Z^2 \sin 2 \beta \over 32 \pi^2 v^2} 
\left \{ {m_{\tau}^2 A_{\tau} \Delta_{{\tilde \tau}_1} \over \cos^2 \beta}
+ {\Delta_{\tilde \tau} \over 2} \right \}
{\log (m_{{\tilde \tau}_2}^2 / m_{{\tilde \tau}_1}^2)
 \over (m_{{\tilde \tau}_2}^2 - m_{{\tilde \tau}_1}^2)} \cr
& & \cr
& &\mbox{} + {m_Z^2 \sin 2 \beta \over 128 \pi^2 v^2}
\left ({4 m_{\tau}^2 \over \cos^2 \beta} - m_Z^2 \right)
\log \left ({m_{{\tilde \tau}_1}^2 m_{{\tilde \tau}_2}^2 \over \Lambda^4} \right ) \ , \cr
& & \cr
M_{13}^{\tau} & = & \mbox{} - {m_{\tau}^2 \mu A_{\tau} \sin \phi_{\tau} \over 4 \pi^2 v^2 \cos^2 \beta}
\left \{{m_{\tau}^2 A_b \Delta_{{\tilde \tau}_1} \over \cos \beta}
+ {\cos \beta \Delta_{\tilde \tau} \over 2} \right \}
{f_2 (m_{{\tilde \tau}_1}^2, \ m_{{\tilde \tau}_2}^2) \over
(m_{{\tilde \tau}_2}^2 - m_{{\tilde \tau}_1}^2)^2 }  \cr
& & \cr
& &\mbox{} + {m_{\tau}^2 \mu A_{\tau} \sin \phi_{\tau} \over 16 \pi^2 v^2 \cos \beta}
\left ( {4 m_{\tau}^2 \over \cos^2 \beta} - m_Z^2 \right )
{ \log (m_{{\tilde \tau}_2}^2 / m_{{\tilde \tau}_1}^2) \over
(m_{{\tilde \tau}_2}^2 - m_{{\tilde \tau}_1}^2)}  \ , \cr
& & \cr
M_{23}^{\tau} & = &\mbox{} - {m_{\tau}^2 \mu A_{\tau} \sin \phi_{\tau}
\over 4 \pi^2 v^2 \cos^2 \beta}
\left \{{m_{\tau}^2 \mu \Delta_{{\tilde \tau}_2} \over \cos \beta}
- {\sin \beta \Delta_{\tilde \tau} \over 2} \right \}
{f_2 (m_{{\tilde \tau}_1}^2, \ m_{{\tilde \tau}_2}^2) \over
(m_{{\tilde \tau}_2}^2 - m_{{\tilde \tau}_1}^2)^2 }  \cr
& & \cr
& &\mbox{} + {m_{\tau}^2 m_Z^2 \mu A_{\tau} \tan \beta \sin \phi_{\tau} \over 16 \pi^2 v^2 \cos \beta}
{\log (m_{{\tilde \tau}_2}^2 / m_{{\tilde \tau}_1}^2)
\over (m_{{\tilde \tau}_2}^2 - m_{{\tilde \tau}_1}^2)} \ , 
\end{eqnarray}
with
\begin{eqnarray}
\Delta_{{\tilde \tau}_1} & = & A_{\tau} - \mu \tan \beta \cos \phi_{\tau} \ , \cr
\Delta_{{\tilde \tau}_2} & = & \mu \tan \beta - A_{\tau} \cos \phi_{\tau} \ , \cr
\Delta_{\tilde \tau} & = & \left ({3 \over 4} m_Z^2 - m_W^2 \right) 
\left \{ m_L^2 - m_E^2 + \left ({3 \over 4} m_Z^2 - m_W^2 \right) \cos 2 \beta  \right \}  \ .
\end{eqnarray}
\vskip 0.2 in
\noindent
{\large {\bf Appendix F}}
\vskip 0.2 in
\noindent
The elements for the mass  matrix of the neutral Higgs bosons due to the radiative contributions of the $W$ boson, 
charged Higgs boson, and charginos are 
\begin{eqnarray}
M_{11}^{\tilde \chi} & = & {\cos^2 \beta \over 8 \pi^2 v^2}
(4 m_W^2 M_2 \Delta_{{\tilde \chi}_1} - \Delta_{\tilde \chi} )^2
{f_2 (m_{{\tilde \chi}_1}^2, \ m_{{\tilde \chi}_2}^2)
\over (m_{{\tilde \chi}_2}^2 - m_{{\tilde \chi}_1}^2)^2} 
+ {m_W^4 \cos^2 \beta \over 4 \pi^2 v^2} 
\log \left ({m_W^6 m_{C^+}^2 \over m_{{\tilde \chi}_1}^4 m_{{\tilde \chi}_2}^4} \right )     \cr
& &\mbox{} - {m_W^2 \cos^2 \beta \over 2 \pi^2 v^2} 
{(4 m_W^2 M_2 \Delta_{{\tilde \chi}_1} - \Delta_{\tilde \chi}) \over (m_{{\tilde \chi}_2}^2 - m_{{\tilde \chi}_1}^2)}
\log \left ({m_{{\tilde  \chi}_2}^2 \over m_{{\tilde  \chi}_1}^2} \right ) 
- {m_W^4 \cos^2 \beta \over 2 \pi^2 v^2} f_1 (m_{{\tilde \chi}_1}^2, \ m_{{\tilde \chi}_2}^2) 
  \ ,  \cr
 & & \cr
M_{22}^{\tilde \chi} & = & {\sin^2 \beta \over 8 \pi^2 v^2}
(4 m_W^2 \mu \cot \beta \Delta_{{\tilde \chi}_2} + \Delta_{\tilde \chi})^2 
{f_2 (m_{{\tilde \chi}_1}^2, \ m_{{\tilde \chi}_2}^2) \over (m_{{\tilde \chi}_2}^2 - m_{{\tilde \chi}_1}^2)^2}  
+ {m_W^4 \sin^2 \beta \over 4 \pi^2 v^2} 
\log \left ({m_W^6 m_{C^+}^2 \over m_{{\tilde \chi}_1}^4 m_{{\tilde \chi}_2}^4} \right )  \cr
& &\mbox{} - {m_W^2 \sin^2 \beta \over 2 \pi^2 v^2} 
{(4 m_W^2 \mu \cot \beta \Delta_{{\tilde \chi}_2} + \Delta_{\tilde \chi}) \over (m_{{\tilde \chi}_2}^2 
- m_{{\tilde \chi}_1}^2)}
\log \left ({m_{{\tilde  \chi}_2}^2 \over m_{{\tilde  \chi}_1}^2} \right ) 
- {m_W^4 \sin^2 \beta \over 2 \pi^2 v^2} f_1 (m_{{\tilde \chi}_1}^2, \ m_{{\tilde \chi}_2}^2)   \ , \cr
 & & \cr
M_{33}^{\tilde \chi} & = & {2 (m_W^2 \mu M_2 \sin \phi_c)^2 \over \pi^2 v^2}
{f_2 (m_{{\tilde \chi}_1}^2, \ m_{{\tilde \chi}_2}^2) \over 
(m_{{\tilde \chi}_2}^2 - m_{{\tilde \chi}_1}^2)^2}    \ , \cr
 & & \cr
M_{12}^{\tilde \chi} & = & {\sin 2 \beta \over 16 \pi^2 v^2}
(4 M_W^2 M_2 \Delta_{{\tilde \chi}_1} - \Delta_{\tilde \chi})
(4 m_W^2 \mu \cot \beta \Delta_{{\tilde \chi}_2} + \Delta_{\tilde \chi}) 
{f_2 (m_{{\tilde \chi}_1}^2, \ m_{{\tilde \chi}_2}^2) \over 
(m_{{\tilde \chi}_2}^2 - m_{{\tilde \chi}_1}^2)^2} \cr
& &\mbox{} - {m_W^4 \cos \beta \over \pi^2 v^2} {(M_2 \sin \beta \Delta_{{\tilde \chi}_1} 
+ \mu \cos \beta \Delta_{{\tilde \chi}_2}) \over (m_{{\tilde \chi}_2} -m_{{\tilde \chi}_2}) }
\log \left ({m_{{\tilde \chi}_2} \over m_{{\tilde \chi}_1}} \right ) \cr 
& &\mbox{} + {m_W^4 \sin 2 \beta \over 8 \pi^2 v^2}
\log \left ( {m_W^6 m_{C^+}^2 \over m_{{\tilde \chi}_1}^4 m_{{\tilde \chi}_2}^4} \right )  
+ {m_W^4 \sin 2 \beta \over 4 \pi^2 v^2} f_1 (m_{{\tilde \chi}_1}^2, \ m_{{\tilde \chi}_2}^2)   \ ,  \cr
 & & \cr
M_{13}^{\tilde \chi} & = & {m_W^2 M_2 \mu \cos \beta \sin \phi_c \over 2 \pi^2 v^2} 
{(4 m_W^2 M_2 \Delta_{{\tilde \chi}_1} - \Delta_{\tilde \chi}) \over (m_{{\tilde \chi}_2}^2- m_{{\tilde \chi}_1}^2)^2}
f_2 (m_{{\tilde \chi}_1}^2, \ m_{{\tilde \chi}_2}^2) \cr
& &\mbox{} - {m_W^4 M_2 \mu \cos \beta \sin \phi_c \over \pi^2 v^2 (m_{{\tilde \chi}_2}^2 
- m_{{\tilde \chi}_1}^2)}
\log \left ({m_{{\tilde \chi}_2}^2 \over  m_{{\tilde \chi}_1}^2} \right ) \ , \cr
 & & \cr
M_{23}^{\tilde \chi} & = & {m_W^2 M_2 \mu \sin \beta \sin \phi_c \over 2 \pi^2 v^2} 
{(4 m_W^2 \mu \cot \beta \Delta_{{\tilde \chi}_2} +\Delta_{\tilde \chi}) \over (m_{{\tilde \chi}_2}^2 
- m_{{\tilde \chi}_1}^2)^2} f_2 (m_{{\tilde \chi}_1}^2, \ m_{{\tilde \chi}_2}^2)  \cr
& &\mbox{} - {m_W^4 M_2 \mu \sin \beta \sin \phi_c \over \pi^2 v^2 (m_{{\tilde  \chi}_2}^2 
- m_{{\tilde \chi}_1}^2)} \log \left ({m_{{\tilde  \chi}_2}^2 \over m_{{\tilde \chi}_1}^2} \right )   \ .
\end{eqnarray}
with
\begin{eqnarray}
\Delta_{{\tilde \chi}_1} & = & M_2 +\mu \tan \beta \cos \phi_c \ , \cr
\Delta_{{\tilde \chi}_2} & = & M_2 \cos \phi_c + \mu \tan \beta \ ,  \cr
\Delta_{\tilde \chi} & = & 2 m_W^2 (M_2^2 - \mu^2 - 2 m_W^2 \cos 2 \beta) \ . 
\end{eqnarray}

\vfil\eject

\vfil\eject

{\bf Figure Captions}
\vskip 0.3 in
\noindent
Fig. 1 : The plot of the $(1,3)$- and $(2,3)$-elements of the mass matrix of the neutral Higgs bosons at the one-loop 
level with the contribution of terms of $O(g_i^4)$ $(i = 1, 2)$, $M_{13}^{{\tilde \chi}^0}$ (dot-dashed curve), 
$M_{23}^{{\tilde \chi}^0}$ (dashed curve), $M_{13}$ (dotted curve) and $M_{23}$ (solid curve), as a function 
of $\phi_1$ for  $\phi_t = \phi_b = \phi_{\tau} = \phi_c (\phi_2) = \pi/3$, $\Lambda$ = 300 GeV, and ${\bar m}_A$ = 300 GeV,$m_Q$ = 800 GeV, $m_T$ = 400 GeV, $A_t$ = 200 GeV, and $M_2$ = 400 GeV.
We set $\tan \beta$ = 5 and $\mu$ = $-400$ GeV. 
These values for the parameters are the same as Table 5, except that $\phi_1$ is taken as a variable. 
\vfil\eject

{\bf Table Caption}
\begin{table}[ht]
\caption{The elements of the symmetric mass matrix of the neutral Higgs bosons in the $(h_1, h_2, h_3)$-basis, at the 
one-loop level with CP violation for $\phi_t = \phi_b = \phi_{\tau} = \phi_c (\phi_2) = \phi_1 = \pi/3$.
Here, the contributions of terms of $O(g^4_i)$ $(i = 1, 2)$ and $A^0$ are neglected.
The unit is (GeV)$^2$.
The values of the relevant parameters are $\Lambda$ = 300 GeV, ${\bar m}_A$ = 300 GeV, $m_Q$ = 800 GeV, $m_T$ = 400 GeV, $A_t$ = 200 GeV, 
and $M_2$ = 400 GeV.
We set $\tan \beta$ = 5 and $\mu$ = $- 400$ GeV. 
The number in the first row in each column is the tree-level value.
The number in the last row in each column is the sum of all numbers in the preceding rows, representing the value at 
the one-loop level.
The numbers in each column in between the two rows represent the various loop contributions as decomposed in 
the one-loop Higgs potential.}
\begin{center}
\begin{tabular}{c|c|c|c|c|c|c} 
\hline 
\hline
$(i, j)$  & (1, 1) & (2, 2) & (3, 3) & (1, 2) & (1, 3) & (2, 3) \\ 
\hline
\hline
$M_{ij}^0$ &  86856.9 &  11424.0  & 90000.0 & $-18900.1$  &  0.0 &  0.0  \\
\hline
$M_{ij}^t$ &  $-7.2$   & 6362.6  &  $-6.9$  &  229.3 & 7.0  &  $-225.0$  \\
\hline
$M_{ij}^b$ & 0.1   & $-0.008$   &  $-0.001$  & 0.01 & $-0.007$  & 0.003 \\
\hline
$M_{ij}^{\tau}$ &  0.002 & $-0.0001$  &  $-0.00001$ & 0.0002   &  $-0.00009$ & 0.00004 \\
\hline
$M_{ij}^{\tilde \chi}$ &   3.0   & $-268.3$   &  16.5      &  $-31.5$  &   17.1 &40.7 \\
\hline
$M_{ij}^h$ &  $-3.7$   & $-150.4$   &  0.0       &  $-12.3$  &   0.0  & 0.0 \\
\hline
$M_{ij}^{{\tilde \chi}^0}$ &   0.1   & $-146.1$   &  12.4      &  $-17.6$  &  10.2  &31.3 \\
\hline
$M_{ij}$ & 86849.3 &  17221.8 &  90022.0   &  $-18732.3$ & 34.4 & $-152.9$ \\
\hline
\hline
\end{tabular}
\end{center}
\caption{The same as Table 1, except for $\mu$ = $400$ GeV.}
\begin{center}
\begin{tabular}{c|c|c|c|c|c|c} 
\hline 
\hline
$(i, j)$  & (1, 1) & (2, 2) & (3, 3) & (1, 2) & (1, 3) & (2, 3) \\ 
\hline
\hline
$M_{ij}^0$ &  86856.9  & 11424.0 & 90000.0 & $-18900.1$  &  0.0 &  0.0  \\
\hline
$M_{ij}^t$ &  $-0.08$   &  6267.7  & $-6.9$  & $-25.5$  & 0.7 & 226.0  \\
\hline
$M_{ij}^b$ &   0.1    &  $-0.006$ & $-0.001$ & 0.01 & 0.009 & $-0.003$  \\
\hline
$M_{ij}^{\tau}$ &   0.002  &  $-0.00008$ & $-0.00001$ & 0.0002 &   0.0001 & $-0.00003$ \\
\hline
$M_{ij}^{\tilde \chi}$ &  $-14.6$   &  $-288.7$ & 16.8 &  $-77.9$  &  1.7 &  $-37.1$ \\
\hline
$M_{ij}^h$ &   $-3.6$    &  $-147.2$ & 0.0  &  $-13.2$  &  0.0 &  0.0   \\
\hline
$M_{ij}^{{\tilde \chi}^0}$ & $-9.3$ &  $-155.1$ & 12.7 &  $-42.1$  &  $-1.9$ & $-29.9$  \\
\hline
$M_{ij}$ &  86829.3  & 17100.5 & 90022.6 & $-19059.0$ &  0.6 & 158.9  \\
\hline
\hline
\end{tabular}
\end{center}
\end{table}
\begin{table}[ht]
\caption{The same as Table 1, except for $\tan \beta$ = 30.}
\begin{center}
\begin{tabular}{c|c|c|c|c|c|c} 
\hline 
\hline
$(i, j)$  & (1, 1) & (2, 2) & (3, 3) & (1, 2) & (1, 3) & (2, 3) \\ 
\hline
\hline
$M_{ij}^0$ & 89909.3 &  8371.6 & 90000.0 & $-3272.3$ & 0.0 & 0.0 \\
\hline
$M_{ij}^t$ &   $-2.7$  &  6088.2 & $-6.4$  & 139.1 & 4.1 & $-212.8$ \\
\hline
$M_{ij}^b$ &  7.2 &  $-9.3$  & $-1.7$  & 1.7  & $-0.7$ & 4.0  \\
\hline
$M_{ij}^{\tau}$ &     0.1 &  $-0.1$  & $-0.02$ & 0.02 & $-0.009$ & 0.05 \\
\hline
$M_{ij}^{\tilde \chi}$ &   6.7 &  $-279.2$ & 16.7 & 13.7   & 10.9  & 40.0  \\
\hline
$M_{ij}^h$ &  $-0.1$ & $-170.2$ &  0.0 & $-1.6$ &  0.0 &  0.0  \\
\hline
$M_{ij}^{{\tilde \chi}^0}$ &   2.1 &  $-155.7$  &  12.5 & 7.1  &  5.3 &  31.3   \\
\hline
$M_{ij}$ &  89922.7 &  13845.2 & 90021.1 & $-3112.2$ & 19.7 & $-137.3$  \\
\hline
\hline
\end{tabular}
\end{center}
\caption{The same as Table 2, except for $\tan \beta$ = 30.}
\begin{center}
\begin{tabular}{c|c|c|c|c|c|c} 
\hline 
\hline
$(i, j)$  & (1, 1) & (2, 2) & (3, 3) & (1, 2) & (1, 3) & (2, 3) \\ 
\hline
\hline
$M_{ij}^0$ & 89909.3  &   8371.6  &  90000.0 & $-3272.3$ &  0.0 & 0.0  \\
\hline
$M_{ij}^t$ & $-1.6$ & 6073.0 & $-6.4$ & $-106.5$ & 3.2 & 212.9  \\
\hline
$M_{ij}^b$ &  3.2  &  $-9.0$ &  $-1.7$  & 6.2 & 2.7  & $-3.9$  \\
\hline
$M_{ij}^{\tau}$ &   0.05 & $-0.1$ & $-0.02$  & 0.08 & 0.03  & $-0.05$  \\
\hline
$M_{ij}^{\tilde \chi}$ &  3.7 & $-309.5$ & 16.7 &  $-33.0$ &  8.3 & $-39.4$  \\
\hline
$M_{ij}^h$ &   $-0.1$ & $-136.4$ & 0.0  &  $-3.0$  &  0.0 & 0.0  \\
\hline
$M_{ij}^{{\tilde \chi}^0}$ &  0.5 & $-157.3$ & 12.6 &  $-17.4$ & 3.2 & $-31.1$  \\
\hline
$M_{ij}$ &  89915.1 &  13832.2  & 90021.2 &  $-3426.1$ & 17.5 &  138.4  \\
\hline
\hline
\end{tabular}
\end{center}
\end{table}
\begin{table}[ht]
\caption{The elements of the symmetric mass matrix of the neutral Higgs bosons in the $(h_1, h_2, h_3)$-basis, at 
the one-loop level with the contributions of terms of $O(g_i^4)$ $(i = 1, 2)$ and $A^0$.
The setting of the parameters is the same as Table 1.}
\begin{center}
\begin{tabular}{c|c|c|c|c|c|c} 
\hline 
\hline
$(i, j)$  & (1, 1) & (2, 2) & (3, 3) & (1, 2) & (1, 3) & (2, 3) \\ 
\hline
\hline
$M_{ij}^0$ & 86856.9 & 11424.0 &  90000.0 & $-18900.1$ & 0.0 & 0.0 \\
\hline
$M_{ij}^t$ &  $-0.7$ & 5832.3 & $-7.0$ & 265.7 & 4.3 &  $-212.3$  \\
\hline
$M_{ij}^b$ &  1.0  & 30.6 & $-0.001$ & $-5.1$ & 0.02 & $-0.1$ \\
\hline
$M_{ij}^{\tau}$ &   0.1 & 4.7  & $-0.00001$ & $-0.8$ & 0.002 & $-0.01$ \\
\hline
$M_{ij}^{\tilde \chi}$ &   1.7 & $-311.8$  & 16.6 & $-23.8$ & 17.3 &  40.0  \\
\hline
$M_{ij}^h$ &  53.1 & $-148.6$ & 55.9 & $-31.5$ &  0.0 & 0.0  \\
\hline
$M_{ij}^{{\tilde \chi}^0}$ &  2.8 & $-82.3$ & 12.4 & $-15.5$ & 10.5 & 31.8  \\
\hline
$M_{ij}$ &  86915.1 &  16749.0 &  90069.7 & $-18711.4$ & 32.2  & $-140.6$ \\
\hline
\hline
\end{tabular}
\end{center}
\caption{The same as Table 5, except for $\mu$ = $400$ GeV.}
\begin{center}
\begin{tabular}{c|c|c|c|c|c|c} 
\hline 
\hline
$(i, j)$  & (1, 1) & (2, 2) & (3, 3) & (1, 2) & (1, 3) & (2, 3) \\ 
\hline
\hline
$M_{ij}^0$ &  86856.9 & 11424.0 & 90000.0 &  $-18900.1$ & 0.0 & 0.0 \\
\hline
$M_{ij}^t$ &  0.1 & 5744.0 & $-6.9$ & 24.7  & 3.5 & 213.3 \\
\hline
$M_{ij}^b$ &  1.0 & 30.5 & $-0.001$ & $-5.3$ & $-0.02$ & 0.1  \\
\hline
$M_{ij}^{\tau}$ & 0.1 & 4.7 & $-0.00001$ & $-0.9$ & $-0.002$ & 0.01  \\
\hline
$M_{ij}^{\tilde \chi}$ &  $-16.2$ & $-331.7$ & 16.9 & $-69.6$ & 1.6 & $-36.4$  \\
\hline
$M_{ij}^h$ &  53.9 & $-146.1$ & 58.0 & $-31.8$ & 0.0 &  0.0  \\
\hline
$M_{ij}^{{\tilde \chi}^0}$ &  $-6.7$ & $-95.5$ & 13.0 & $-46.3$ &  0.5 & $-30.1$  \\
\hline
$M_{ij}$ &  86889.3 &  16630.0 & 90072.5 & $-19029.5$ & 5.6 & 146.9  \\
\hline
\hline
\end{tabular}
\end{center}
\end{table}
\begin{table}[ht]
\caption{The same as Table 5, except for $\tan \beta$ = 30.}
\begin{center}
\begin{tabular}{c|c|c|c|c|c|c} 
\hline 
\hline
$(i, j)$  & (1, 1) & (2, 2) & (3, 3) & (1, 2) & (1, 3) & (2, 3) \\ 
\hline
\hline
$M_{ij}^0$ &  89909.3 & 8371.6 & 90000.0 & $-3272.3$ & 0.0  & 0.0 \\
\hline
$M_{ij}^t$ &   $-2.1$ & 5561.2 & $-6.4$ & 139.2 & 3.7 & $-200.4$ \\
\hline
$M_{ij}^b$ &  6.7 &  51.3 & $-1.7$ & 8.4 & $-0.5$ & $-2.4$  \\
\hline
$M_{ij}^{\tau}$ &   0.07 & 7.0 &  $-0.02$ & 0.3 & 0.006 & $-0.4$  \\ 
\hline
$M_{ij}^{\tilde \chi}$ &  6.7 &  $-324.4$ & 16.7 & 14.7 & 10.9 &  39.2 \\
\hline
$M_{ij}^h$ &  40.0 & $-168.3$ & 30.2 & $-4.0$ & 0.0 & 0.0 \\
\hline
$M_{ij}^{{\tilde \chi}^0}$ &  5.0 & $-91.3$ & 12.8 & 12.8  & 8.1 & 32.1 \\ 
\hline
$M_{ij}$ &  89965.7 &  13407.2 & 90043.2 & $-3100.9$ & 22.4 & $-131.8$ \\ 
\hline
\hline
\end{tabular}
\end{center}
\caption{The same as Table 6, except for $\tan \beta$ = 30.}
\begin{center}
\begin{tabular}{c|c|c|c|c|c|c} 
\hline 
\hline
$(i, j)$  & (1, 1) & (2, 2) & (3, 3) & (1, 2) & (1, 3) & (2, 3) \\ 
\hline
\hline
$M_{ij}^0$ &  89909.3  &  8371.6 & 90000.0 & $-3272.3$  & 0.0  &  0.0  \\
\hline
$M_{ij}^t$ &   $-2.0$  &  5547.0 & $-6.4$ & $-92.1$ & 3.6 &  200.5  \\
\hline
$M_{ij}^b$ &  3.2  &  51.1  &  $-1.7$  &  5.5   & 2.4  & 2.4  \\
\hline
$M_{ij}^{\tau}$ &  0.06  &  6.9  &  $-0.02$ &  $-0.1$ & 0.01 &  0.4  \\
\hline
$M_{ij}^{\tilde \chi}$ &  3.6  &  $-354.7$ & 16.8 &  $-31.2$ &  8.3 &  $-38.6$ \\
\hline
$M_{ij}^h$ &  38.9  & $-140.8$ &  39.2 &  $-3.8$ &  0.0 &    0.0  \\
\hline
$M_{ij}^{{\tilde \chi}^0}$ &  2.5  &  $-93.6$ & 12.9  &  $-22.7$ & 6.0 &  $-31.7$  \\
\hline
$M_{ij}$ &  89955.7 & 13387.7 & 90052.3 &  $-3417.0$  & 20.5 & 133.0 \\
\hline
\hline
\end{tabular}
\end{center}
\end{table}

\vfil\eject

\setcounter{figure}{0}
\def\figurename{}{}%
\renewcommand\thefigure{Fig. 1}
\begin{figure}[t]
\epsfxsize=13cm
\hspace*{2.cm}
\epsffile{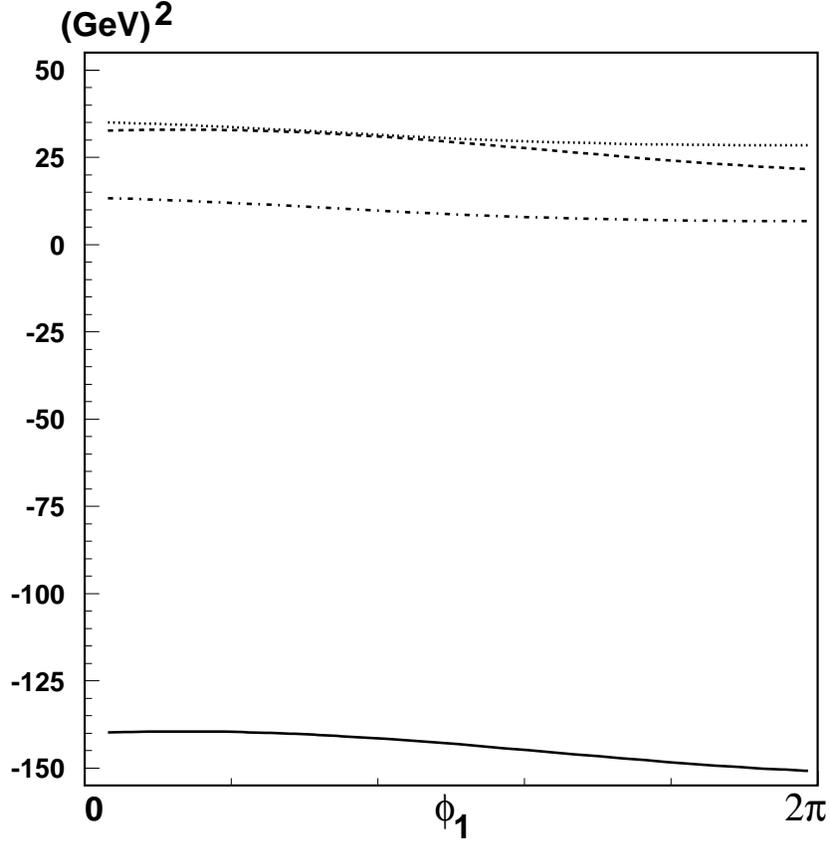}
\caption[plot]{The plot of the $(1,3)$- and $(2,3)$-elements of the mass matrix of the neutral Higgs bosons at the 
one-loop level with the contribution of the quartic terms of $O(g_i^4)$ $(i = 1, 2)$, $M_{13}^{{\tilde \chi}^0}$ (dot-dashed curve), 
$M_{23}^{{\tilde \chi}^0}$ (dashed curve), $M_{13}$ (dotted curve) and $M_{23}$ (solid curve), as a 
function of $\phi_1$ for  $\phi_t = \phi_b = \phi_{\tau} = \phi_c (\phi_2) = \pi/3$, $\Lambda$ = 300 GeV, and ${\bar m}_A$ = 300 GeV, $m_Q$ = 800 GeV, $m_T$ = 400 GeV, $A_t$ = 200 GeV, and $M_2$ = 400 GeV.
We set $\tan \beta$ = 5 and $\mu$ = $-400$ GeV. 
These values for the parameters are the same as Table 5, except that $\phi_1$ is taken as a variable.}
\end{figure}

\end{document}